\documentclass[zdraft,zbstdraft,american]{./zeus_K2.9paper}
\usepackage{graphicx}
\usepackage{epstopdf}
\usepackage{pdflscape}

\usepackage{./zeusdet_units}
\chardef\usc=95
\chardef\til=126
\catcode`\@=11 
\DeclareRobustCommand\xdotspace{\futurelet\@let@token\@xdotspace}
\def\@xdotspace{%
  \ifx\@let@token.\else
  \ifx\@let@token\bgroup.\else
  \ifx\@let@token\egroup.\else
  \ifx\@let@token\/.\else
  \ifx\@let@token\ .\else
  \ifx\@let@token~.\else
  \ifx\@let@token!.\else
  \ifx\@let@token,.\else
  \ifx\@let@token:.\else
  \ifx\@let@token;.\else
  \ifx\@let@token?.\else
  \ifx\@let@token/.\else
  \ifx\@let@token'.\else
  \ifx\@let@token).\else
  \ifx\@let@token-.\else
  \ifx\@let@token\@xobeysp.\else
  \ifx\@let@token\space.\else
  \ifx\@let@token\@sptoken.\else
   .\space
   \fi\fi\fi\fi\fi\fi\fi\fi\fi\fi\fi\fi\fi\fi\fi\fi\fi\fi}
\catcode`\@=12 

\newcommand{\stru}[2]{%
   \relax\ifmmode\hbox{\vrule height#1 depth#2 width0pt}%
   \else\vrule height#1 depth#2 width0pt\fi}

\newcommand{\Ronum}[1]{\uppercase\expandafter{\romannumeral#1}}
\newcommand{\ronum}[1]{\expandafter{\romannumeral#1}}
\DeclareRobustCommand{\LaTeXZ}{%
  \LaTeX\kern-.05em4\kern-.1em
  {\raisebox{-0.2ex}{$\scriptstyle\text{ZEUS}$}}\xspace}

\newcommand{\fig}[1]{Fig.~\ref{fig-#1}}

\newcommand{\slashfrac}[2]{%
  \raisebox{0.5ex}{\ensuremath #1}\kern-0.12em/\kern-0.08em
  \raisebox{-.8ex}{\ensuremath #2}}

\newcommand{\sqr}[3]{%
    {\vcenter{\hrule height.#3ex\hbox{\vrule width.#2ex height#1ex
     \kern#1ex\vrule width.#3ex}\hrule height.#2ex}}}

\catcode`\@=11 
\newcommand{\parenbar}{\mathpalette\p@renb@r}
\def\p@renb@r#1#2{\vbox{%
  \ifx#1\scriptscriptstyle \dimen@.7em\dimen@ii.2em\else
  \ifx#1\scriptstyle \dimen@.8em\dimen@ii.25em\else
  \dimen@1em\dimen@ii.4em\fi\fi \offinterlineskip
  \ialign{\hfill##\hfill\cr
    \vbox{\hrule width\dimen@ii}\cr
    \noalign{\vskip-.3ex}%
    \hbox to\dimen@{$\mathchar300\hfil\mathchar301$}\cr
    \noalign{\vskip-.3ex}%
    $#1#2$\cr}}}
\catcode`\@=12 

\ifzeusunit
  
\else
  
\fi

\newcommand{\IP}{{\rm I$\kern-0.01667em$P}\xspace}

\mathchardef\qsm=63
\mathchardef\pls=43
\mathchardef\mns=512
\mathchardef\plm=518
\mathchardef\eql=61
\mathchardef\smallleft=300
\mathchardef\smallright=301
\mathchardef\les=316
\mathchardef\gre=318
\mathchardef\leq=532
\mathchardef\grq=533

\catcode`\@=11 
\newcounter{pict@width}
\newcounter{pict@height}
\newlength{\pict@scale}
\newcommand{\psfigadd}[4]{%
\setcounter{pict@width}{1*\ratio{#2+\pict@scale/2}{\pict@scale}}
\setcounter{pict@height}{1*\ratio{#3+\pict@scale/2}{\pict@scale}}
\setlength{\unitlength}{\pict@scale}
\hbox to #2{\hspace{-\fill}\begin{picture}(\thepict@width,\thepict@height)
\put(0,0){\psfig{figure=#1,width=#2,height=#3,clip=}}
\SetScale{0.283466457}
\SetWidth{1.763889}
{#4}
\end{picture}}
}
\newcounter{pict@widthfst}
\newcounter{pict@widthscd}
\newcounter{pict@widthtot}
\newcommand{\psfigaddtwo}[7]{%
\setcounter{pict@widthfst}{1*\ratio{#2+\pict@scale/2}{\pict@scale}}
\setcounter{pict@widthscd}{1*\ratio{#2+#4+\pict@scale/2}{\pict@scale}}
\setcounter{pict@widthtot}{1*\ratio{#2+#4+#6+\pict@scale/2}{\pict@scale}}
\setcounter{pict@height}{1*\ratio{#3+\pict@scale/2}{\pict@scale}}
\setlength{\unitlength}{\pict@scale}
\hbox{\hspace{-\fill}\begin{picture}(\thepict@widthtot,\thepict@height)
\put(0,0){\psfig{figure=#1,width=#2,height=#3,clip=}}
\put(\thepict@widthscd,0){\psfig{figure=#5,width=#6,height=#3,clip=}}
\SetScale{0.283466457}
\SetWidth{1.763889}
{#7}
\end{picture}}
}
\newcommand{\psfigror}[4]{%
\setcounter{pict@width}{1*\ratio{#2+\pict@scale/2}{\pict@scale}}
\setcounter{pict@height}{1*\ratio{#3+\pict@scale/2}{\pict@scale}}
\setlength{\unitlength}{\pict@scale}
\hbox{\begin{picture}(\thepict@width,\thepict@height)
\put(0,\thepict@height){\psfig{figure=#1,width=#3,height=#2,clip=,angle=270}}
\SetScale{0.283466457}
\SetWidth{1.763889}
{#4}
\end{picture}}
}
\newcommand{\psfigrol}[4]{%
\setcounter{pict@width}{1*\ratio{#2+\pict@scale/2}{\pict@scale}}
\setcounter{pict@height}{1*\ratio{#3+\pict@scale/2}{\pict@scale}}
\setlength{\unitlength}{\pict@scale}
\hbox{\begin{picture}(\thepict@width,\thepict@height)
\put(0,0){\psfig{figure=#1,width=#3,height=#2,clip=,angle=90}}
\SetScale{0.283466457}
\SetWidth{1.763889}
{#4}
\end{picture}}
}
\catcode`\@=12 
\newlength\listtextwidth

\catcode`\@=11 
\newlength{\@tabfninsert}
\newlength{\@tabfnwidth}
\newcommand{\tabfootnote}[2]{%
  \setlength{\@tabfninsert}{0.8em}
  \setlength{\@tabfnwidth}{\textwidth}
  \addtolength{\@tabfnwidth}{-\@tabfninsert}
  \addtolength{\@tabfnwidth}{-0.4em}
  \noindent\makebox[\@tabfninsert][r]{\footnotesize$^{#1}$\hfil}\hfill%
  \parbox[t]{\@tabfnwidth}{\footnotesize #2\hfill}}
\catcode`\@=12 

\begin{document}


\zeustitle{%
\vspace{-5cm}
\begin{flushright}
{\normalsize \tt DESY--16--008}\\
\vspace{-.25cm}{\normalsize \tt January 2016}
\end{flushright}
\vspace{2cm}
Measurement of the cross-section ratio {\boldmath $\sigma_{\psi(2S )}/\sigma_{J/\psi(1S )}$}
in deep inelastic exclusive {\boldmath $ep$} scattering at HERA
}

\zeusauthor{ZEUS Collaboration}

 \zeusdate{}

\maketitle

%
%
\begin{abstract}\noindent
 The exclusive deep inelastic electroproduction of $\psi(2S)$ and $J/\psi(1S)$ at an $ep$ centre-of-mass energy of 317\,GeV has been studied with the ZEUS detector at HERA in the kinematic range $2 < Q^2 < 80$\,GeV$^2$, $30 < W < 210$\,GeV and $|t| < 1$\,GeV$^2$, where $Q^2$ is the photon virtuality, $W$ is the photon--proton centre-of-mass energy and $t$ is the squared four-momentum transfer at the proton vertex.
 The data for $2 < Q^2 < 5$\,GeV$^2$ were taken in
the HERA\,I running period and correspond to an integrated luminosity of 114\,pb$^{-1}$.
 The data for $5 < Q^2 < 80$\,GeV$^2$ are from both HERA\,I and HERA\,II periods and
correspond to an integrated luminosity of 468\,pb$^{-1}$.
 The decay modes analysed were $\mu^+\mu^-$ and $J/\psi(1S) \,\pi^+\pi^-$ for the $\psi(2S)$ and $\mu^+\mu^-$ for the $J/\psi(1S)$.
 The cross-section ratio $\sigma_{\psi(2S)}/\sigma_{J/\psi(1S)}$ has been measured as a function of $Q^2,\,W$\, and $t$.
 The results are compared to predictions of QCD-inspired models of exclusive vector-meson production.

\end{abstract}

  \thispagestyle{empty}
\clearpage

%
%
%
%


\topmargin-1.cm
\evensidemargin-0.3cm
\oddsidemargin-0.3cm
\textwidth 16.cm
\textheight 680pt
\parindent0.cm
\parskip0.3cm plus0.05cm minus0.05cm
\def\3{\ss}
\pagenumbering{Roman}
                                                   %
\begin{center}
{                      \Large  The ZEUS Collaboration              }
\end{center}

{\small\raggedright


H.~Abramowicz$^{26, v}$, 
I.~Abt$^{21}$, 
L.~Adamczyk$^{8}$, 
M.~Adamus$^{32}$, 
S.~Antonelli$^{2}$, 
V.~Aushev$^{16, 17, p}$, 
Y.~Aushev$^{17, p, q}$, 
O.~Behnke$^{10}$, 
U.~Behrens$^{10}$, 
A.~Bertolin$^{23}$
% ...
\newpage


{\setlength{\parskip}{0.4em}
\makebox[3ex]{$^{1}$}
\begin{minipage}[t]{14cm}
{\it INFN Bologna, Bologna, Italy}~$^{A}$

\end{minipage}

\makebox[3ex]{$^{2}$}
\begin{minipage}[t]{14cm}
{\it University and INFN Bologna, Bologna, Italy}~$^{A}$

\end{minipage}

\makebox[3ex]{$^{3}$}
\begin{minipage}[t]{14cm}
{\it Physikalisches Institut der Universit\"at Bonn,
Bonn, Germany}~$^{B}$

\end{minipage}

\makebox[3ex]{$^{4}$}
\begin{minipage}[t]{14cm}
{\it Panjab University, Department of Physics, Chandigarh, India}

\end{minipage}

\makebox[3ex]{$^{5}$}
\begin{minipage}[t]{14cm}
{\it Calabria University,
Physics Department and INFN, Cosenza, Italy}~$^{A}$

\end{minipage}

\makebox[3ex]{$^{6}$}
\begin{minipage}[t]{14cm}
{\it National Centre for Particle Physics, Universiti Malaya, 50603 Kuala Lumpur, Malaysia}~$^{C}$

\end{minipage}

\makebox[3ex]{$^{7}$}
\begin{minipage}[t]{14cm}
{\it The Henryk Niewodniczanski Institute of Nuclear Physics, Polish Academy of \\
Sciences, Krakow, Poland}~$^{D}$

\end{minipage}

\makebox[3ex]{$^{8}$}
\begin{minipage}[t]{14cm}
{\it AGH-University of Science and Technology, Faculty of Physics and Applied Computer
Science, Krakow, Poland}~$^{D}$

\end{minipage}

\makebox[3ex]{$^{9}$}
\begin{minipage}[t]{14cm}
{\it Department of Physics, Jagellonian University, Krakow, Poland}

\end{minipage}

\makebox[3ex]{$^{10}$}
\begin{minipage}[t]{14cm}
{\it Deutsches Elektronen-Synchrotron DESY, Hamburg, Germany}

\end{minipage}

\makebox[3ex]{$^{11}$}
\begin{minipage}[t]{14cm}
{\it Deutsches Elektronen-Synchrotron DESY, Zeuthen, Germany}

\end{minipage}

\makebox[3ex]{$^{12}$}
\begin{minipage}[t]{14cm}
{\it School of Physics and Astronomy, University of Glasgow,
Glasgow, United Kingdom}~$^{E}$

\end{minipage}

\makebox[3ex]{$^{13}$}
\begin{minipage}[t]{14cm}
{\it Hamburg University, Institute of Experimental Physics, Hamburg,
Germany}~$^{F}$

\end{minipage}

\makebox[3ex]{$^{14}$}
\begin{minipage}[t]{14cm}
{\it Institute of Particle and Nuclear Studies, KEK,
Tsukuba, Japan}~$^{G}$

\end{minipage}

\makebox[3ex]{$^{15}$}
\begin{minipage}[t]{14cm}
{\it Institute of Physics and Technology of Ministry of Education and
Science of Kazakhstan, Almaty, Kazakhstan}

\end{minipage}

\makebox[3ex]{$^{16}$}
\begin{minipage}[t]{14cm}
{\it Institute for Nuclear Research, National Academy of Sciences, Kyiv, Ukraine}

\end{minipage}

\makebox[3ex]{$^{17}$}
\begin{minipage}[t]{14cm}
{\it Department of Nuclear Physics, National Taras Shevchenko University of Kyiv, Kyiv, Ukraine}

\end{minipage}

\makebox[3ex]{$^{18}$}
\begin{minipage}[t]{14cm}
{\it Department of Physics, McGill University,
Montr\'eal, Qu\'ebec, Canada H3A 2T8}~$^{H}$

\end{minipage}

\makebox[3ex]{$^{19}$}
\begin{minipage}[t]{14cm}
{\it Meiji Gakuin University, Faculty of General Education,
Yokohama, Japan}~$^{G}$

\end{minipage}

\makebox[3ex]{$^{20}$}
\begin{minipage}[t]{14cm}
{\it Lomonosov Moscow State University, Skobeltsyn Institute of Nuclear Physics,
Moscow, Russia}~$^{I}$

\end{minipage}

\makebox[3ex]{$^{21}$}
\begin{minipage}[t]{14cm}
{\it Max-Planck-Institut f\"ur Physik, M\"unchen, Germany}

\end{minipage}

\makebox[3ex]{$^{22}$}
\begin{minipage}[t]{14cm}
{\it Department of Physics, University of Oxford,
Oxford, United Kingdom}~$^{E}$

\end{minipage}

\makebox[3ex]{$^{23}$}
\begin{minipage}[t]{14cm}
{\it INFN Padova, Padova, Italy}~$^{A}$

\end{minipage}

\makebox[3ex]{$^{24}$}
\begin{minipage}[t]{14cm}
{\it Dipartimento di Fisica e Astronomia dell' Universit\`a and INFN,
Padova, Italy}~$^{A}$

\end{minipage}

\makebox[3ex]{$^{25}$}
\begin{minipage}[t]{14cm}
{\it Polytechnic University, Tokyo, Japan}~$^{G}$

\end{minipage}

\makebox[3ex]{$^{26}$}
\begin{minipage}[t]{14cm}
{\it Raymond and Beverly Sackler Faculty of Exact Sciences, School of Physics, \\
Tel Aviv University, Tel Aviv, Israel}~$^{J}$

\end{minipage}

\makebox[3ex]{$^{27}$}
\begin{minipage}[t]{14cm}
{\it Department of Physics, Tokyo Institute of Technology,
Tokyo, Japan}~$^{G}$

\end{minipage}

\makebox[3ex]{$^{28}$}
\begin{minipage}[t]{14cm}
{\it Universit\`a di Torino and INFN, Torino, Italy}~$^{A}$

\end{minipage}

\makebox[3ex]{$^{29}$}
\begin{minipage}[t]{14cm}
{\it Universit\`a del Piemonte Orientale, Novara, and INFN, Torino,
Italy}~$^{A}$

\end{minipage}

\makebox[3ex]{$^{30}$}
\begin{minipage}[t]{14cm}
{\it Physics and Astronomy Department, University College London,
London, United Kingdom}~$^{E}$

\end{minipage}

\makebox[3ex]{$^{31}$}
\begin{minipage}[t]{14cm}
{\it Faculty of Physics, University of Warsaw, Warsaw, Poland}

\end{minipage}

\makebox[3ex]{$^{32}$}
\begin{minipage}[t]{14cm}
{\it National Centre for Nuclear Research, Warsaw, Poland}

\end{minipage}

\makebox[3ex]{$^{33}$}
\begin{minipage}[t]{14cm}
{\it Department of Particle Physics and Astrophysics, Weizmann
Institute, Rehovot, Israel}

\end{minipage}

\makebox[3ex]{$^{34}$}
\begin{minipage}[t]{14cm}
{\it Department of Physics, York University, Ontario, Canada M3J 1P3}~$^{H}$

\end{minipage}

}

\vspace{3em}


{\setlength{\parskip}{0.4em}\raggedright
\makebox[3ex]{$^{ A}$}
\begin{minipage}[t]{14cm}
 supported by the Italian National Institute for Nuclear Physics (INFN) \
\end{minipage}

\makebox[3ex]{$^{ B}$}
\begin{minipage}[t]{14cm}
 supported by the German Federal Ministry for Education and Research (BMBF), under
 contract No. 05 H09PDF\
\end{minipage}

\makebox[3ex]{$^{ C}$}
\begin{minipage}[t]{14cm}
 supported by HIR grant UM.C/625/1/HIR/149 and UMRG grants RU006-2013, RP012A-13AFR and RP012B-13AFR from
 Universiti Malaya, and ERGS grant ER004-2012A from the Ministry of Education, Malaysia\
\end{minipage}

\makebox[3ex]{$^{ D}$}
\begin{minipage}[t]{14cm}
 supported by the National Science Centre under contract No. DEC-2012/06/M/ST2/00428\
\end{minipage}

\makebox[3ex]{$^{ E}$}
\begin{minipage}[t]{14cm}
 supported by the Science and Technology Facilities Council, UK\
\end{minipage}

\makebox[3ex]{$^{ F}$}
\begin{minipage}[t]{14cm}
 supported by the German Federal Ministry for Education and Research (BMBF), under
 contract No. 05h09GUF, and the SFB 676 of the Deutsche Forschungsgemeinschaft (DFG) \
\end{minipage}

\makebox[3ex]{$^{ G}$}
\begin{minipage}[t]{14cm}
 supported by the Japanese Ministry of Education, Culture, Sports, Science and Technology
 (MEXT) and its grants for Scientific Research\
\end{minipage}

\makebox[3ex]{$^{ H}$}
\begin{minipage}[t]{14cm}
 supported by the Natural Sciences and Engineering Research Council of Canada (NSERC) \
\end{minipage}

\makebox[3ex]{$^{ I}$}
\begin{minipage}[t]{14cm}
 supported by RF Presidential grant N 3042.2014.2 for the Leading Scientific Schools and by
 the Russian Ministry of Education and Science through its grant for Scientific Research on
 High Energy Physics\
\end{minipage}

\makebox[3ex]{$^{ J}$}
\begin{minipage}[t]{14cm}
 supported by the Israel Science Foundation\
\end{minipage}

}

\pagebreak[4]
{\setlength{\parskip}{0.4em}


\makebox[3ex]{$^{ a}$}
\begin{minipage}[t]{14cm}
also funded by Max Planck Institute for Physics, Munich, Germany, now at Sant Longowal Institute of Engineering and Technology, Longowal, Punjab, India\
\end{minipage}

\makebox[3ex]{$^{ b}$}
\begin{minipage}[t]{14cm}
also funded by Max Planck Institute for Physics, Munich, Germany, now at Sri Guru Granth Sahib World University, Fatehgarh Sahib, India\
\end{minipage}

\makebox[3ex]{$^{ c}$}
\begin{minipage}[t]{14cm}
also at Agensi Nuklear Malaysia, 43000 Kajang, Bangi, Malaysia\
\end{minipage}

\makebox[3ex]{$^{ d}$}
\begin{minipage}[t]{14cm}
now at School of Physics and Astronomy, University of Birmingham, UK\
\end{minipage}

\makebox[3ex]{$^{ e}$}
\begin{minipage}[t]{14cm}
partially supported by the Polish National Science Centre projects DEC-2011/01/B/ST2/03643 and DEC-2011/03/B/ST2/00220\
\end{minipage}

\makebox[3ex]{$^{ f}$}
\begin{minipage}[t]{14cm}
now at Rockefeller University, New York, NY 10065, USA\
\end{minipage}

\makebox[3ex]{$^{ g}$}
\begin{minipage}[t]{14cm}
now at European X-ray Free-Electron Laser facility GmbH, Hamburg, Germany\
\end{minipage}

\makebox[3ex]{$^{ h}$}
\begin{minipage}[t]{14cm}
now at University of Liverpool, United Kingdom\
\end{minipage}

\makebox[3ex]{$^{ i}$}
\begin{minipage}[t]{14cm}
now at Physikalisches Institut, Universit\"{a}t Heidelberg, Germany\
\end{minipage}

\makebox[3ex]{$^{ j}$}
\begin{minipage}[t]{14cm}
supported by the Alexander von Humboldt Foundation\
\end{minipage}

\makebox[3ex]{$^{ k}$}
\begin{minipage}[t]{14cm}
now at CERN, Geneva, Switzerland\
\end{minipage}

\makebox[3ex]{$^{ l}$}
\begin{minipage}[t]{14cm}
Alexander von Humboldt Professor; also at DESY and University of Oxford\
\end{minipage}

\makebox[3ex]{$^{ m}$}
\begin{minipage}[t]{14cm}
also at DESY\
\end{minipage}

\makebox[3ex]{$^{ n}$}
\begin{minipage}[t]{14cm}
also at University of Tokyo, Japan\
\end{minipage}

\makebox[3ex]{$^{ o}$}
\begin{minipage}[t]{14cm}
now at Kobe University, Japan\
\end{minipage}

\makebox[3ex]{$^{ p}$}
\begin{minipage}[t]{14cm}
supported by DESY, Germany\
\end{minipage}

\makebox[3ex]{$^{ q}$}
\begin{minipage}[t]{14cm}
member of National Technical University of Ukraine, Kyiv Polytechnic Institute, Kyiv, Ukraine\
\end{minipage}

\makebox[3ex]{$^{ r}$}
\begin{minipage}[t]{14cm}
now at DESY ATLAS group\
\end{minipage}

\makebox[3ex]{$^{ s}$}
\begin{minipage}[t]{14cm}
member of National University of Kyiv - Mohyla Academy, Kyiv, Ukraine\
\end{minipage}

\makebox[3ex]{$^{ t}$}
\begin{minipage}[t]{14cm}
now at BNL, USA\
\end{minipage}

\makebox[3ex]{$^{ u}$}
\begin{minipage}[t]{14cm}
now at LNF, Frascati, Italy\
\end{minipage}

\makebox[3ex]{$^{ v}$}
\begin{minipage}[t]{14cm}
also at Max Planck Institute for Physics, Munich, Germany, External Scientific Member\
\end{minipage}

\makebox[3ex]{$^{ w}$}
\begin{minipage}[t]{14cm}
also at Universit\"{a}t Hamburg and supported by DESY and the Alexander von Humboldt Foundation\
\end{minipage}

\makebox[3ex]{$^{ x}$}
\begin{minipage}[t]{14cm}
also at \L\'{o}d\'{z} University, Poland\
\end{minipage}

\makebox[3ex]{$^{ y}$}
\begin{minipage}[t]{14cm}
member of \L\'{o}d\'{z} University, Poland\
\end{minipage}

\makebox[3ex]{$^{ z}$}
\begin{minipage}[t]{14cm}
now at Polish Air Force Academy in Deblin\
\end{minipage}

}

}


\cleardoublepage
\pagenumbering{arabic}
%
%
\pagestyle{scrheadings}
 
\section{Introduction}
\label{sec-int}

\begin{figure}[p]
\vfill
\begin{center}
\includegraphics[width=12.0cm]{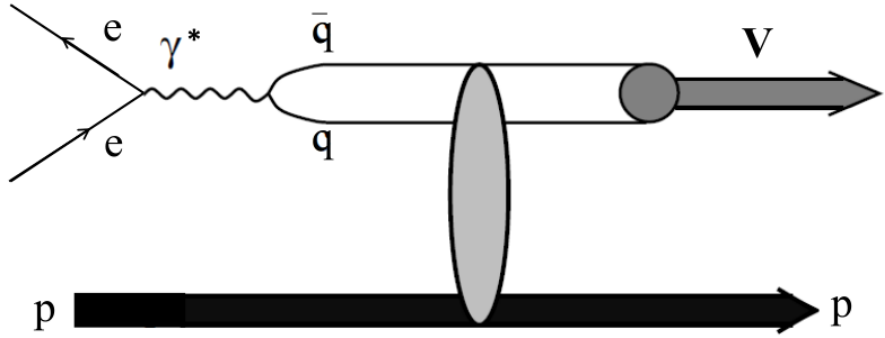}
\end{center}
 \caption{Schematic representation of the exclusive electroproduction of $q \bar{q}$~vector mesons.
 The electron emits a virtual photon, which fluctuates into a $q \bar{q} $~pair. The $q \bar{q} $~pair interacts with the target proton and produces the $q \bar{q}$~vector meson $V$.}
\label{fig-diagram}
\vfill
\end{figure}

 Exclusive electroproduction of vector mesons in deep inelastic scattering (DIS) at high energies is usually described as a multi-step process, as illustrated in Fig.~\ref{fig-diagram}:
 the electron emits a virtual photon, $\gamma^*$, with virtuality, $Q^2$;
 the $\gamma^*$ fluctuates in leading-order QCD into a $q\bar{q}$ pair with a lifetime which, at large values of the $\gamma^* p$ centre-of-mass energy, $W$, is long compared to the interaction time;
 and  the $q \bar{q}$ pair interacts with the proton with momentum transfer squared, $t$, via a colour-neutral exchange, e.g.\ through a two-gluon ladder, and then hadronises into the vector meson, $V$.

 In this paper, a measurement of the ratio of the cross sections of the reactions $\gamma ^* p \rightarrow \psi(2S ) + Y$ and  $\gamma ^* p \rightarrow J/\psi(1S ) + Y$, where $Y$ denotes either a proton or a low-mass proton-dissociative system, is presented.
The $\psi (2S)$ and the $J/\psi (1S)$ have the same quark content, different radial distributions of the wave functions, and their mass difference is small compared to the HERA centre-of-mass energy.
 Therefore, this measurement allows QCD predictions of the wave function dependence of the $c\bar{c}$--proton cross section to be tested.
 A suppression of the $\psi (2S)$ cross section relative to the $J/\psi(1S)$ is expected, as the $\psi(2S)$~wave function has a radial node close to the typical transverse separation of the virtual $c \bar{c}$\, pair, as will be discussed in more detail in Section\,\ref{sect:Models}.

 At HERA, deep inelastic exclusive $J/\psi(1S)$ electroproduction has been measured for $2 < Q^2 < 100 $\,GeV$^2 $, $30 < W < 220 $\,GeV and $ |t| < 1$\,GeV$^2$ by the ZEUS collaboration~\cite{Chekanov:2004mw} and for $2 < Q^2 < 80 $\,GeV$^2 $, $25 < W < 180 $\,GeV and $|t| < 1.6$\,GeV$^2$  by the H1 collaboration~\cite{epj:c10:373}.
 The H1 collaboration has also measured the quasi-elastic component, $\gamma ^* p \rightarrow \psi(2S ) + Y$, in  DIS~\cite{epj:c10:373} and photoproduction~\cite{pl:b421:385}, as well as the ratio of the $\psi(2S)$ to $J/\psi(1S)$ production cross sections.

 The luminosity used for the measurement of
 $\sigma (e p \to e\,\psi(2S)\,Y) / \sigma (e p \to e\,J/\psi(1S)\,Y) $
 presented in this paper is 468\,pb$^{-1}$ and the kinematic range is $5 <Q^2 <80$\,GeV$^2$, $30 <W <210$\,GeV and $|t| <1$\,GeV$^2$.
 A sub-sample of 114\,pb$^{-1}$ of HERA\,I data was used for an additional measurement in the range $2 <Q^2 <5$\,GeV$^2$.
 Events were selected with no activity in the central ZEUS detector in addition to signals from the scattered electron and the decay products of the $\psi(2S)$ or $J/\psi(1S)$.
 The decay channels used were $J/\psi(1S) \to \mu^+ \mu^-$, $\psi(2S)\to \mu^+\mu^-$ and $\psi(2S)\,\to J/\psi(1S)\,\pi^+\pi^-$ with the subsequent decay $J/\psi(1S )\to \mu^+\mu^-$.

\newpage

 \section{Experimental set-up}
  \label{sec-exp}

  The measurement is based on data collected with the ZEUS detector at the HERA collider during the period 1996--2007 where an electron\footnote{Hereafter electron refers to both electrons and positrons unless otherwise stated.} beam of energy 27.5\,GeV collided with a proton beam of either 820\,GeV (1996--97) or 920\,GeV (1998--2007).
  The integrated lumi\-nosity was 38\,pb$^{-1}$ and 430\,pb$^{-1}$ for $ep$ centre-of-mass energies of 300\,GeV and 318\,GeV, respectively.
  The luminosity-weighted $ep$ centre-of-mass energy is 317\,GeV.

 A detailed description of the ZEUS detector can be found elsewhere~\cite{zeus:1993:bluebook}. A brief outline of the components that are most relevant for this analysis is given below.


 In the kinematic range of the analysis, charged particles were tracked in the central tracking detector (CTD)~\cite{nim:a279:290,npps:b32:181,nim:a338:254}, which operated in a magnetic field of 1.43\,T provided by a thin superconducting solenoid.
 The CTD consisted of 72~cylindrical drift-chamber layers, organised in nine superlayers covering the polar-angle\footnote{The ZEUS coordinate system is a right-handed Cartesian system, with the $Z$ axis pointing in the proton beam direction, referred to as the ``forward direction'', and the $X$ axis pointing towards the centre of HERA.
  The coordinate origin is at the nominal beam-crossing point\,\cite{zeus:1993:bluebook}.}
region 15$^\circ < \theta < 164 ^\circ $.
 The transverse-momentum resolution for full-length tracks was $\sigma(p_{T})/p_{T} = 0.0058 p_{T} \oplus 0.0065 \oplus 0.0014/p_{T}$\,, with $p_{T}$ in\,GeV.
 For the HERA\,II period, the CTD was complemented by a silicon microvertex detector (MVD)~\cite{nim:a581:656}, which consisted of three cylindrical layers of silicon microstrip sensors in the central region and four planar disks in the forward region.

  The high-resolution uranium--scintillator calorimeter
(CAL)~\cite{nim:a309:77, nim:a309:101, nim:a321:356, nim:a336:23} consisted of three parts:
the forward (FCAL), the barrel (BCAL) and the rear (RCAL) calorimeters. Each
part was subdivided transversely into towers and longitudinally into one
electromagnetic section (EMC) and either one (in RCAL) or two (in BCAL and FCAL)
hadronic sections (HAC). The CAL energy reso\-lutions, as measured under test-beam conditions,
were $\sigma(E)/E=0.18/\sqrt{E}$ for electrons and $\sigma(E)/E=0.35/\sqrt{E}$
for hadrons, with $E$ in GeV.

The muon system consisted of rear, barrel (R/BMUON) and forward
(FMUON) tracking detectors. The R/BMUON consisted of limited-streamer
(LS) tube chambers placed behind the RCAL (BCAL), inside and outside a
magnetised iron yoke surrounding the CAL. The barrel and rear muon
chambers covered polar angles from 34$^\circ$ to 135$^\circ$ and from
135$^\circ$ to 171$^\circ$, respectively. The FMUON consisted of six
trigger planes of LS tubes and four planes of drift chambers covering
the angular region from 5$^\circ$ to 32$^\circ$. The muon system
exploited the magnetic field of the iron yoke and, in the forward
direction, of two iron toroids magnetised to $\approx 1.6$\,T
to provide an independent measurement of the muon momenta.

 The iron yoke surrounding the CAL was instrumented with proportional drift chambers to form the backing calorimeter (BAC)~\cite{nim:a300:480}.
 The BAC consisted of 5142 aluminium chambers inserted into the gaps between $7.3\,{\rm cm}$ thick iron plates (10, 9 and 7 layers in the forward, barrel and rear directions, respectively).
 The chambers were typically $5\,{\rm m}$ long and had a wire spacing of $1.5\,{\rm cm}$.
 The anode wires were covered by $50\,{\rm cm}$ long cathode pads.
 The BAC was equipped with energy readout and position-sensitive readout for muon tracking.
 The former was based on 1692 pad towers ($50 \times 50\,{\rm cm^2}$), providing an energy resolution of $\sigma(E)/E \approx 100\%/\sqrt E$, with $E$ in GeV.
 The position information from the wires allowed the reconstruction of muon trajectories in two dimensions ($XY$ in the barrel and $YZ$ in the endcaps) with a spatial accuracy of a few mm.

 The luminosity was measured using the Bethe--Heitler reaction $ep\,\rightarrow\, e\gamma p$ by a luminosity detector which consisted of a lead--scintillator calorimeter~\cite{desy-92-066, zfp:c63:391, Andruszkow:2001jy} and, additionally in HERA\,II, a magnetic spectrometer\cite{nim:a565:572}.

 \section{Monte Carlo simulations}
  \label{sec-dataset}

 The {\sc Diffvm}~\cite{proc:mc:1998:396} Monte Carlo (MC) programme was used for simulating exclusive vector meson production, $ep \rightarrow e V p$, where $V$ denotes the produced vector meson with mass $M_V$.
 For the event generation, the following cross-section parameterisations were used:
 $(1 + Q^2/M_V^2)^{-1.5}$ for transverse virtual photons;
 $(Q/M_V)^{2}$ for the cross-section ratio of longitudinal to transverse photons;
 $\exp(-b\,|t|)$, with $b = 4$\,GeV$^{-2}$ for the dependence on $t$;
 $s$-channel helicity conservation for the production of $V \rightarrow \mu ^+ \mu ^-$;
 and a flat angular distribution for the $\psi (2S) \to J/\psi (1S ) \pi ^+ \pi ^-$ decay.
 The relative contributions of the proton-dissociative part to the cross sections of the $\psi (2S)$ and $J/\psi (1S)$  is expected to be similar and due
 to factorisation~\cite{Chekanov:2007zr}, cancel in the cross-section ratio.
 Therefore it was not necessary to simulate proton-dissociative events.
 Radiative effects were not simulated.
 The largest contribution is expected to come from the initial-state radiation of the electrons, however, as the $\psi(2S) - J/\psi(1S)$\,mass difference is small compared to $W$, the kinematics of both reactions are similar and radiative effects are expected to cancel in the cross-section ratio.

 Non-resonant electroweak dimuon production (Bethe--Heitler background) was simulated using the dimuon programme {\sc Grape}~\cite{cpc:136:126}.
 The event sample contains both exclusive and proton-dissociative events with a mass of the dissociated proton system, $M_Y < 25$\,GeV.

 The generated MC events were passed through the ZEUS detector and trigger simulation programmes based on {\sc Geant} 3~\cite{tech:cern-dd-ee-84-1}.
 They were then reconstructed and analysed with the same programmes as used for the data.

\section{Event selection and signal extraction}
  \label{sec-event_selection}

\subsection{Event selection}
 A three-level trigger system~\cite{zeus:1993:bluebook, nim:a580:1257, desy-92-150b} was  used to select events online.
 For this analysis, DIS events were selected with triggers containing a candidate scattered electron.
 Further triggers were used with a less stringent electron-candidate selection in coincidence with a muon candidate or with two tracks.

 The offline event selection required an electron candidate in the RCAL with an energy
 $E_{e}' >10\,\rm{GeV}$ as reconstructed using an algorithm based on a neural network~\cite{nim:a365:508}.
 The position of the scattered electron was required to be outside areas with significant inactive material in front of the calorimeter.

 To select events containing exclusively produced $J/\psi(1S)$ and $\psi(2S)$ vector mesons, the following additional requirements were imposed:

\begin{itemize}

\item the $Z$ coordinate of the interaction vertex reconstructed from the tracks was required to be within $\pm 30$\,cm of
the nominal interaction point;

\item in addition to the scattered electron, two oppositely charged muons were required.
    Muons were identified using the GMUON algorithm~\cite{thesis:bloch:2005} with muon quality $\ge 1$.
    This algorithm required a track with momentum above 1\,GeV to be matched with a cluster in the CAL.
    The cluster was required to be consistent with a muon using an algorithm based on a neural network~\cite{nim:a453:336};

\item for selecting $J/\psi(1S)\,\to\,\mu^+\mu^-$ and $\psi(2S)\,\to\,\mu^+\mu^-$ events, no additional tracks were allowed;

\item for selecting $\psi(2S)\,\to\,J/\psi(1S )\,\pi^+\pi^-$~events, exactly two oppositely charged tracks were required in addition to the two muons.
    The tracks had to cross at least three CTD superlayers, produce hits in the first CTD superlayer or in the MVD and the transverse momentum of each track had to exceed 0.12\,GeV;

\item events with calorimeter clusters with energies above $0.4\,\rm{GeV}$, not associated with the electron or the decay products of the vector meson, were rejected.
    The threshold value of this cut was optimised by minimising event loss from calorimeter noise fluctuations and maximising the rejection of non-exclusive vector-meson production with additional energy deposits in the CAL.
\newline

\end{itemize}
 The last three requirements significantly reduced the background from non-exclusive charmonium production and also removed proton-dissociative events with diffractive masses $M_Y \gtrsim 4$~GeV~\cite{cpc:136:126}.

\subsection{Reconstruction of the kinematic variables and signal extraction}

 The kinematic variables, $Q^2$, $W$ and $t$ were used in the analysis. The constrained method~\cite{epj:c6:603} was used to reconstruct $Q^2$ according to
  $$Q^2 = 2E_{e}\, E^{\prime}_{e}(1 + \cos\theta_{e})\,, $$
 where
  $$E^{\prime}_{e} = \frac{2E_{e} - (E - p_{Z})_{V}}{1 - \cos\theta_{e}}\,.$$
 Here $E_{e}$ denotes the electron beam energy and $E_{e}^{\prime}$ and $\theta_{e}$ the energy and the polar angle of the scattered electron, respectively.
 The quantity $(E - p_{Z})_{V}$ denotes the difference of the energy and the $Z$ component of the momentum of $V$.
 The momentum components of the vector-meson candidate, $V$, and its effective mass, $M_V$, were obtained from the momentum vectors of the decay products measured by the tracking detectors.
 The values of $W$ and $t$ were calculated using
   $$ W = \sqrt{2\,E_p \,(E-p_Z)_V - Q^2 + M_p^2}\,$$
 and
   $$ -t = (\vec{p}_{T,e}^{\,\,\prime} + \vec{p}_{T,V})^2\,. $$
 Here $E_p$ is the proton beam energy, $M_p$ is the proton mass and $\vec{p}_{T,e}^{\,\,\prime}$ and $\vec{p}_{T,V}$ are the transverse-momentum vectors of the scattered electron and of the vector-meson candidate, respectively.
 The kinematic range for the analysis of the HERA\,II (HERA\,I) data is $5\,(2) <Q^{2} <80\,\rm{GeV^2}$, $30 < W <210\,\rm{GeV}$ and $|t| <1\,\rm{GeV^2}$.
 The range of Bjorken\,$x$, $x_\mathrm{Bj} \approx (M_V^2 + Q^2)/(W^2 + Q^2)\,,$ probed by the measurements is $2\times 10^{-4} < x_\mathrm{Bj} < 10^{-2}$.

\begin{figure}[p]
\vfill
\begin{center}
\includegraphics[width=15.cm]{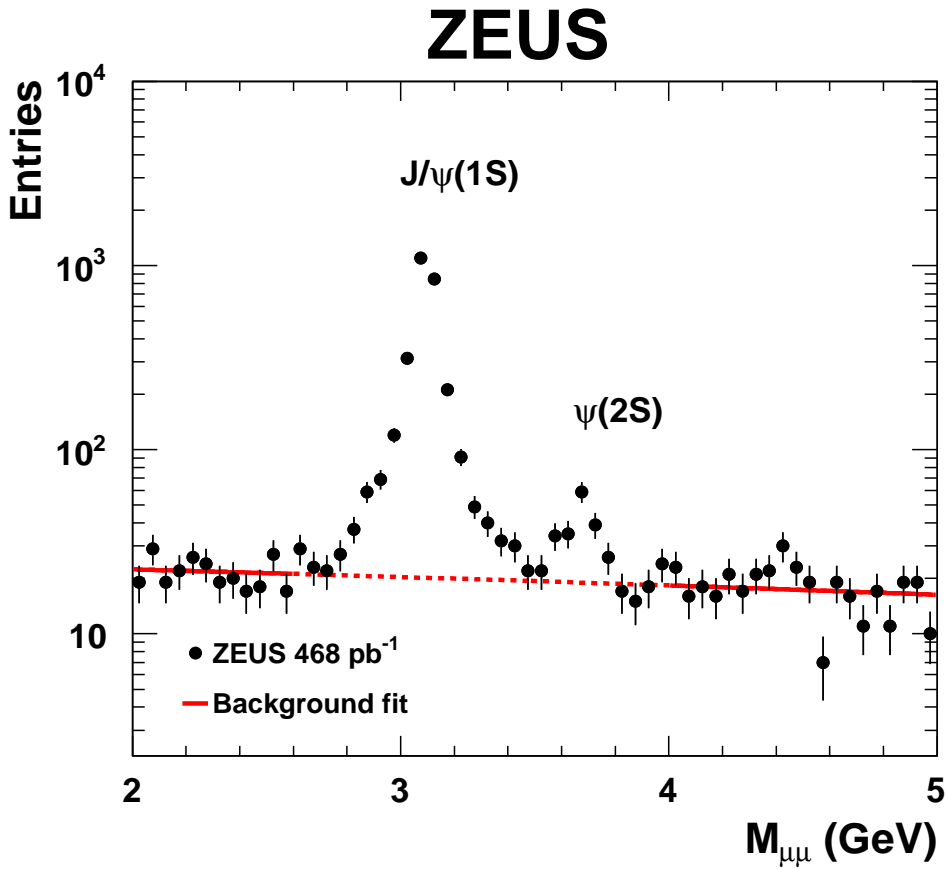}
\end{center}
 \caption{Two-muon invariant-mass distribution, $M_{\mu\mu}$, for exclusive dimuon events. The data (points) are shown with statistical uncertainties.  The background distribution (solid line) is described by a linear fit to the data outside of the $J/\psi(1S)$ and $\psi(2S)$ signal regions, and is also shown (dashed line) in the signal regions.}
\label{fig-dimuon-mass}
\vfill
\end{figure}

 Figure~\ref{fig-dimuon-mass} shows the $\mu ^+ \mu ^-$ mass distribution for the selected events in the region $5 <Q^{2} <80\,\rm{GeV^2}$.
 Clear $J/\psi (1S)$ and $\psi(2S)$ peaks are seen.
 No other significant peak  is observed.
 The background was fit by a straight line in the side-bands of the signals: $2.0 < M_{\mu \mu} < 2.62$\,GeV and $4.05 < M_{\mu \mu} < 5.0 $\,GeV.

 The ratio of $\psi (2S)$ to $J/\psi(1S)$ events for the $\mu^+\mu^-$ decay channel was obtained from the ratio of the number of events above background in the range $ 3.59 < M_{\mu \mu} < 3.79 $\,GeV  to the corresponding number in the range $3.02 < M_{\mu \mu} < 3.17 $\,GeV.
 According to a detailed MC study, this choice minimises the systematic uncertainty due to the uncertainties of the mass-resolution function and the shape of the background.
 The difference in widths of the mass bins chosen takes into account the worsening of the mass resolution with increasing $M_{\mu \mu }$.

 To study the systematic uncertainty related to the background subtraction, a quadratic background function was used and the widths of the $M_{\mu \mu }$ intervals for the signal determination were varied.
 The Bethe--Heitler MC events provide a good description of the background shape and its absolute normalisation after acceptance correction agrees with the measured rate within the estimated uncertainty of about 20\,\%.
 Given this uncertainty, the linear fit described in the previous paragraph was used for the background subtraction.

\begin{figure}[p]
\vfill
\begin{center}
\includegraphics[width=15.0cm]{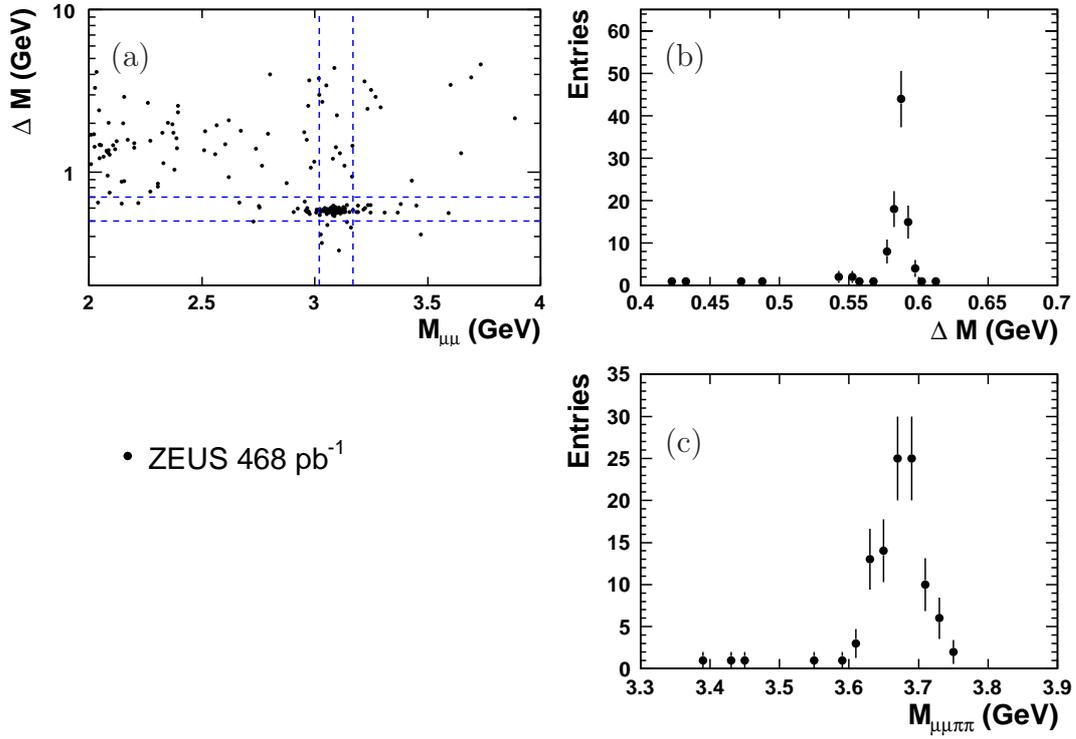}
\\[-95mm]
\hspace*{-40.5mm}(a)\hspace*{67.5mm}(b)
\\[46mm]
\hspace*{32mm}(c)
\\[45mm]

\end{center}
 \caption{ (a) Scatter plot of $\Delta M = M_{\mu \mu \pi \pi} - M_{\mu \mu}$ versus $M_{\mu \mu}$, for the selected $\mu \mu \pi \pi$ events,
   (b)\,$\Delta M $\,distribution for $3.02 < M_{\mu \mu} < 3.17$\,GeV and
   (c)\,$M_{\mu \mu \pi \pi}$\,distribution for $3.02 < M_{\mu \mu} < 3.17$\,GeV.
   }
\label{fig-psi-mass}
\vfill
\end{figure}

 Figure~\ref{fig-psi-mass} shows a scatter plot of $\Delta M = M_{\mu \mu \pi \pi} - M_{\mu \mu}$ versus $M_{\mu \mu}$ and the $\Delta M $\, and $M_{\mu \mu \pi \pi}$ distributions for $3.02 < M_{\mu \mu} < 3.17$\,GeV.
 As expected, the $\Delta M$\,distribution shows a narrow peak with a width of about 5\,MeV at the nominal $\psi(2S ) - J/\psi(1S )$ mass difference of $589.2$\,MeV.
 The mass ranges of $ 3.02 < M_{\mu \mu} < 3.17 $\,GeV and $0.5 < \Delta M < 0.7$\,GeV were chosen to compute the ratio of $\psi (2S)$ to $J/\psi(1S)$ events.
 As can be seen from \fig{psi-mass}, there is hardly any background in the $\psi (2S)$ signal region; an upper limit of three\,events at 90\,\% confidence level was estimated for the background.

 The numbers of events and their statistical uncertainties used for the further analysis for the kinematic region $5 < Q^2 < 80$~GeV$^2$, $ 30 < W < 210$~GeV and $|t| < 1$\,GeV$^2$ are $2224 \pm 48,~97 \pm 10 $ and $80 \pm 13$ for the $J/\psi \to \mu^+ \mu^-$, $\psi(2S)\,\to\,J/\psi(1S )\,\pi^+\pi^-$ and the $\psi(2S) \to \mu^+ \mu^-$ decays, respectively.
 For $2 < Q^2 < 5$~GeV$^2$, the corresponding numbers are $297 \pm 18,~11.0 \pm 3.3 $ and $ 4.4 \pm 4.1$.

\subsection{Comparison of measured and simulated distributions}
\label{sec:data-MC}

 In order to determine the acceptance using simulated events,  simulated and measured distributions have to agree.
 To achieve this, the simulated events had to be  reweighted.
 The reweighting functions for the $J/\psi(1S)$ as well as for the $\psi(2S)$ events were obtained by comparing the measured distributions of the $J/\psi(1S) \to \mu ^+ \mu^- $\,sample to the simulated distributions. For the reweighting of the $t$ and $Q^2$ distributions, a two-dimensional function was used.
 No reweighting was required for the $W$\,distribution.
 The following reweighting function was used for the angular distribution for the vector meson decays into $\mu^+ \mu^-$:
 $$ f(\Phi _h) = 1 - \epsilon \cdot \cos(2 \Phi _h) \cdot (2 r_{11}^1 + r_{00}^1) + \sqrt{ 2\, \epsilon \,(1 + \epsilon)} \cdot \cos (\Phi _h) \cdot ( 2 r_{11}^5 + r_{00}^5)\,. $$
 The helicity angle $\Phi_h$ is the angle between the production and scattering planes, where the production plane is defined by the three-momenta of the photon and the vector meson, and the scattering plane is defined by the three-momenta of the incoming and the scattered electron.
 The elements of the spin-density matrix, $r_{ij}^k$, were obtained by fitting the weighted simulated events to the measured decay angular distribution of $J/\psi(1S)$. The dominant contribution comes from $r_{00}^1$, compatible with $s$-channel helicity conservation.
 The quantity $\epsilon$ denotes the ratio of the longitudinal to the transverse virtual-photon flux, which was set to unity in the kinematic range of the measurement.
 In the MC simulation, the fitted angular distributon was used as reweighting function for both vector-meson decays into $\mu^+ \mu^-$, whereas no reweighting was applied for the flatly simulated decay $\psi(2S) \to J/\psi(1S ) \,\pi ^+ \pi^- $.

\begin{figure}[p]
\vfill
\begin{center}
\includegraphics[width=15.0cm]{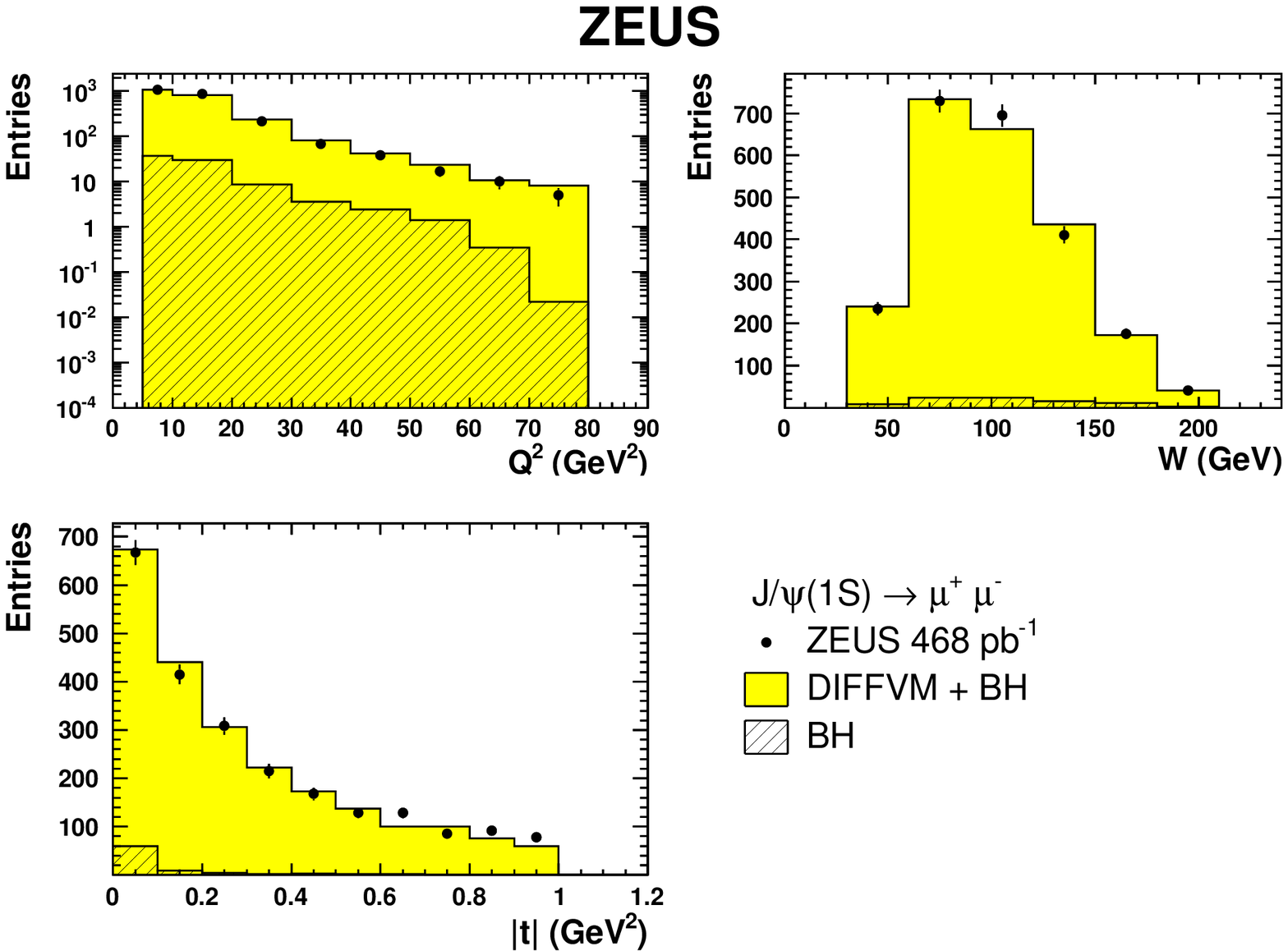}
\\[-95mm]
\hspace*{46.5mm}(a)\hspace*{65.5mm}(b)
\\[45mm]
\hspace*{-25mm}(c)
\\[45mm]
\end{center}
 \caption{
 Comparison of the measured (points with statistical uncertainties) and the reweighted simulated distributions (shaded histograms) for the $J/\psi(1S) \to \mu^+ \mu^- $~events as a function of (a) $Q^2$, (b) $W$ and (c) $|t|$.
 The mass range $3.02 < M_{\mu \mu} < 3.17$\,GeV was  selected.
 The hatched histogram shows the contribution of the simulated Bethe--Heitler background.
 }
\label{fig-controlplots1}
\vfill
\end{figure}
\begin{figure}[p]
\vfill
\begin{center}
\includegraphics[width=15.0cm]{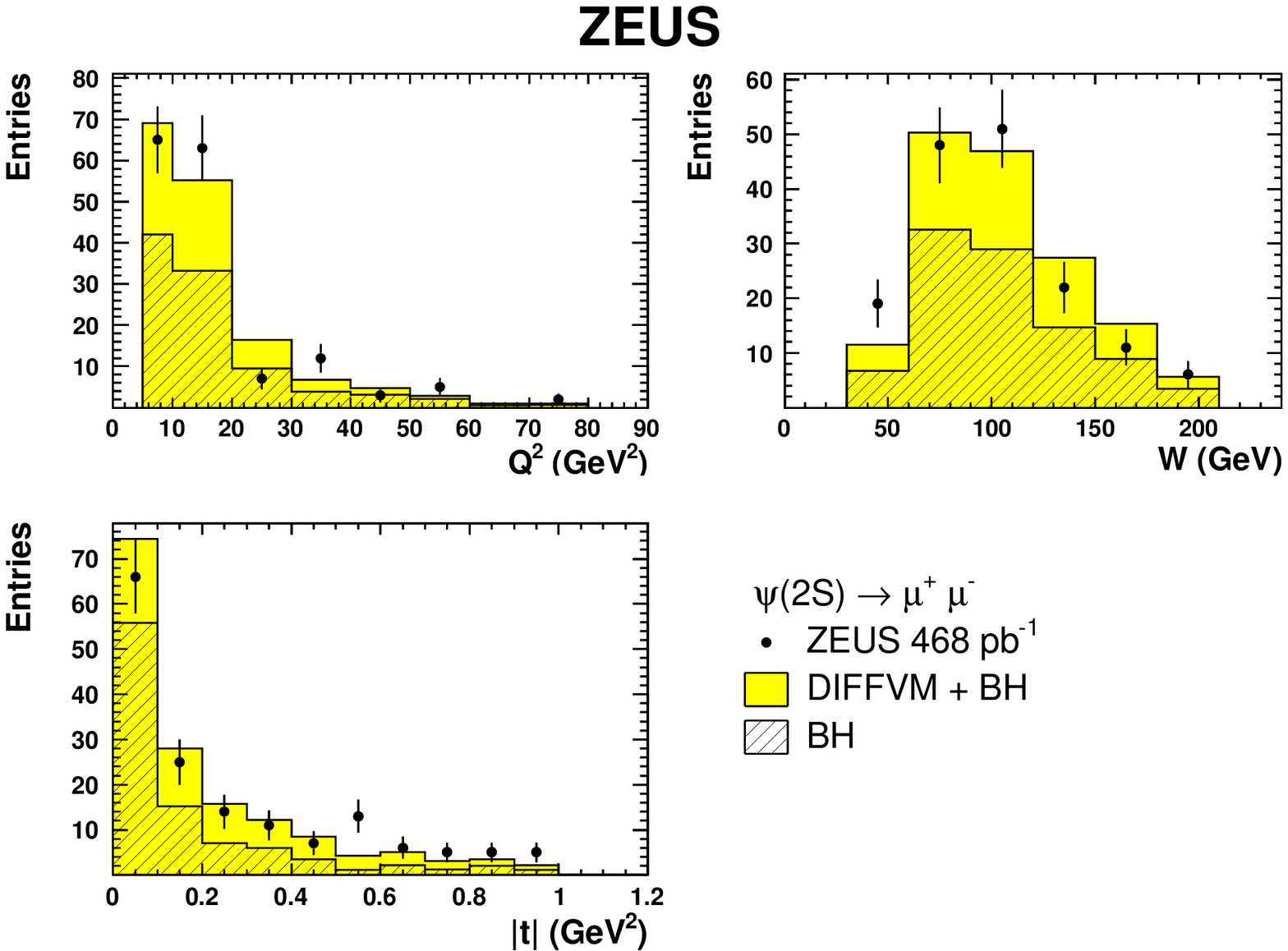}
\\[-95mm]
\hspace*{46.5mm}(a)\hspace*{65.5mm}(b)
\\[45mm]
\hspace*{-25mm}(c)
\\[45mm]
\end{center}
 \caption{
 Comparison of the measured (points with statistical uncertainties) and the reweighted simulated distributions (shaded histograms) for the $\psi(2S) \to \mu^+ \mu^- $~events as a function of (a) $Q^2$, (b) $W$ and (c) $|t|$.
 The mass range $3.59 < M_{\mu \mu} < 3.79$\,GeV was selected.
 The hatched histogram shows the contribution of the simulated Bethe--Heitler background.
 }
\label{fig-controlplots2}
\vfill
\end{figure}
\begin{figure}[p]
\vfill
\begin{center}
\includegraphics[width=15.0cm]{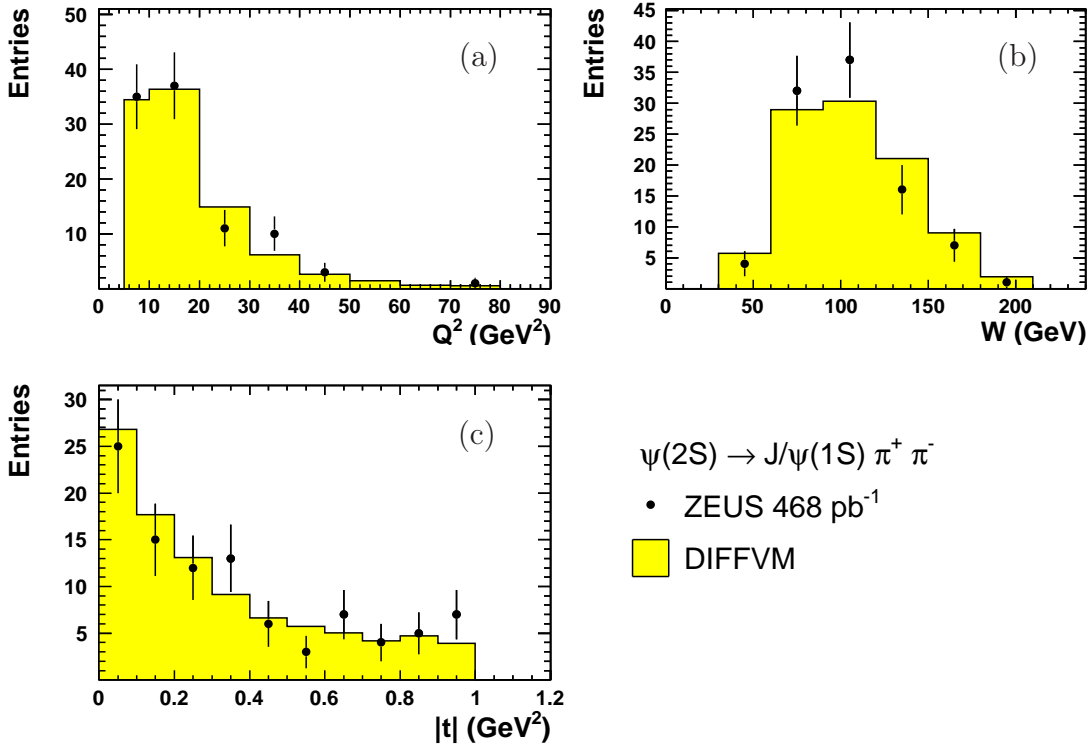}
\\[-95mm]
\hspace*{46.5mm}(a)\hspace*{65.5mm}(b)
\\[45mm]
\hspace*{-25mm}(c)
\\[45mm]
\end{center}
 \caption{
 Comparison of the measured (points with statistical uncertainties) and the reweighted simulated distributions (shaded histograms) for the $\psi(2S) \rightarrow J/\psi(1S)\, \pi^+ \pi^- $~events as a function of (a) $Q^2$, (b) $W$ and (c) $|t|$.
 The mass ranges $ 3.02 < M_{\mu \mu} < 3.17 $\,GeV and $0.5 < \Delta M < 0.7$\,GeV were selected.
 }
\label{fig-controlplots3}
\vfill
\end{figure}

 The comparisons of the $Q^2$, $W$ and $|t|$\,distributions between data and reweighted MC events normalised to the number of measured events are shown in Figs.~\ref{fig-controlplots1},~\ref{fig-controlplots2} and~\ref{fig-controlplots3} for the $J/\psi(1S)\,\to\,\mu^+\mu^-$, $\psi(2S)\,\to\,\mu^+\mu^-$ and $\psi(2S)\,\to\,J/\psi(1S)\,\pi^+\pi^-$ channels, respectively.
 Agreement is observed for all distributions.


\section{Cross-section ratio $\boldsymbol{\psi (2S)}$ to $\boldsymbol{J/\psi (1S)}$}
 \label{ratio}

 The following cross-section ratios, $\sigma_{\psi(2S)} / \sigma_{J/\psi (1S)}$, have been measured:
  $R_{\mu \mu}$ for $\psi (2S ) \rightarrow \mu ^+ \mu^-$,
  $R_{J/\psi \,\pi \pi} $ for $\psi (2S ) \rightarrow J/\psi (1S ) \,\pi^+ \pi^-$ and
  $R_{\rm comb}$ for the combination of the two decay modes.
  In each case the decay $J/\psi (1S) \to \mu ^+ \mu^-$ was used.

\subsection{Determination of the cross-section ratio}
 \label{Ratio determination}

 The cross-section ratios were calculated using
$$ R_{\mu \mu} =
   \left( \frac{N_{\mu \mu}^{\psi (2S)}} {B(\psi (2S) \to \mu ^+ \mu^-) \cdot A_{\mu \mu}^{\psi (2S)} } \right)\Big/
   \left( \frac{N_{\mu \mu}^{J/\psi(1S)}} {B(J/\psi (1S) \to \mu ^+ \mu^-) \cdot A_{\mu \mu}^{J/\psi(1S)}} \right)
   \, \mathrm{and} $$
$$ R_{J/\psi \, \pi \pi} =
   \left(\frac{N_{J/\psi \,\pi \pi}^{\psi (2S)}} {B(\psi (2S ) \to J/\psi (1S ) \,\pi^+ \pi^-) \cdot A_{J/\psi \, \pi \pi}^{\psi (2S)}} \right) \Big/
   \left(\frac{N_{\mu \mu}^{J/\psi(1S)}} {A_{\mu \mu}^{J/\psi(1S)}} \right) \,, $$
 where $N_i^j$ denotes the number of observed signal events for the charmonium state $j$ with the decay mode $i$, and $A_i^j$ the corresponding acceptance determined from the ratio of reconstructed to generated MC events after reweighting.
 The following values were used for the branching fractions:
  $B(J/\psi (1S) \to \mu ^+ \mu^-) = (5.93 \pm 0.06 )\%$,
  $B(\psi (2S) \to \mu ^+ \mu^-) = (0.77 \pm 0.08 )\%$ and
  $B(\psi (2S ) \to J/\psi (1S ) \,\pi^+ \pi^-) = (33.6 \pm 0.4 )\%$\,\cite{Beringer:1900zz}.


 The combined cross-section ratio, $R_{\rm comb}$, was obtained using the weighted average of the cross sections determined for the two $\psi(2S)$ decay modes.
 For the weights, the statistical uncertainties were used.
 The different $R$ values were determined in the full kine\-matic region as well as in bins of $Q^2$, $W$ and $|t|$.
 The results are reported in Section~\ref{sect:Results}.

\subsection{Systematic uncertainties}
    \label{sect:Systematics}

 The systematic uncertainties of the $R$ values were obtained by performing for each source of uncertainty a suitable variation in order to determine the change of $R$ relative to its nominal value.
 The following sources of systematic uncertainties were considered:

\begin{itemize}

\item reducing the $M_{\mu \mu}$ range for the $J/\psi(1S )$ from the nominal value
    $3.02 -3.17 \,\rm{GeV}$ to
    $3.05 -3.15 \,\rm{GeV}$
 and for the $\psi(2S )$ from the nominal value
    $3.59 -3.79 \,\rm{GeV}$ to
    $3.62 -3.75 \,\rm{GeV}$,
 changes the values of $R_{\mu \mu} $ by $\approx $\,+2\% and of $R_{J/\psi \,\pi \pi}$ by $\approx $\,+1.5\%;

\item increasing the $M_{\mu \mu}$ range for the $J/\psi(1S )$ from the nominal value
    $3.02 -3.17 \,\rm{GeV}$ to
    $2.97 -3.22 \,\rm{GeV}$
 and for the $\psi(2S )$ from the nominal value
    $3.59 -3.79 \,\rm{GeV}$ to
    $3.55 -3.80 \,\rm{GeV}$,
 changes the values of $R_{\mu \mu} $ by $\approx $\,6\% and of $R_{J/\psi \,\pi \pi}$ by $\approx $\,$-0.5$\%;

\item changing the cut on the transverse momenta $p_T$ of the pion tracks from the nominal value of $0.12\,\rm{GeV}$ to $0.15\,\rm{GeV}$ changes the values of  $R_{J/\psi \,\pi \pi} $ by $\approx $\,$-4.5$\%;

\item changing the background fit function from linear to quadratic changes the values of $R_{\mu \mu}$ by $ \approx $\,$-11$\% and of $R_{J/\psi \,\pi \pi} $ by $ \approx $\,+0.5\%;

\item changing the reconstruction from the constrained to the electron method~\cite{hoeger} changes the values of $R_{\mu \mu} $ and of $R_{J/\psi \,\pi \pi}$ by $ \approx $\,+1.5\%;


\item not applying the reweighting of the simulated events discussed in Section~\ref{sec:data-MC} changes the values of $R_{\mu \mu}$ by $\approx$\,$-3$\% and of $R_{J/\psi \,\pi \pi}$ by $\approx$\,$-1$\%;



\item applying different cuts on the total number of tracks, including tracks not associated with the event vertex, changes $R_{\mu \mu}$ by $\approx$\,$-$5\% and $R_{J/\psi \,\pi \pi}$ by $\approx$\,+3\%.

 \end{itemize}

The total systematic uncertainty was obtained from the separate quadratic sums of the positive and negative changes.
The estimated total systematic uncertainties are
$ \delta R_{\mu \mu} = \, ^{+7}_{-14}$\,\%,
$ \delta R_{J/\psi \,\pi \pi} = \, ^{+4}_{-5}$\,\% and
$ \delta R_{\rm comb} = \, ^{+3}_{-5}$\,\%.
 For the calculation of $\delta R_{\rm comb}$, the uncertainties of the two measurements were assumed to be uncorrelated.

\section{Results}
      \label{sect:Results}

 The results for the three cross-section ratios
 $\sigma _{\psi(2S )} / \sigma _{J/\psi (1S )}$:
 $R_{\mu \mu}$ for $\psi (2S ) \rightarrow \mu ^+ \mu^-$,
 $R_{J/\psi \,\pi \pi} $ for $\psi (2S ) \rightarrow J/\psi(1S ) \pi^+ \pi^-$ and $R_{\rm comb}$ for the combination,  are reported in Table~\ref{tab:Rtot} for the kinematic range
 $5 < Q^2 < 80$\,GeV$^2$, $30 < W < 210$\,GeV and $|t| < 1$\,GeV$^2$  for the total integrated luminosity of 468~pb$^{-1}$.
 The cross-section ratios, differential in $Q^2$, $W$ and $|t|$, together with the additional measurement between $2 < Q^2 < 5$\,GeV$^2$, corresponding to an integrated luminosity of 114~pb$^{-1}$, are shown in Table~\ref{tab:R-diff}.
 The data contain a  background of charmonium production with diffractive masses $M_Y \lesssim 4$\,GeV.
 Assuming that the  $\psi(2S)$ to $J/\psi(1S)$ cross-section ratio for exclusive charmonium production is the same as for charmonium production with low $M_Y$, the determination of the $R$~values for exclusive production are not affected.

\begin{figure}[p]
\vfill
\begin{center}
\includegraphics[width=8.5cm]{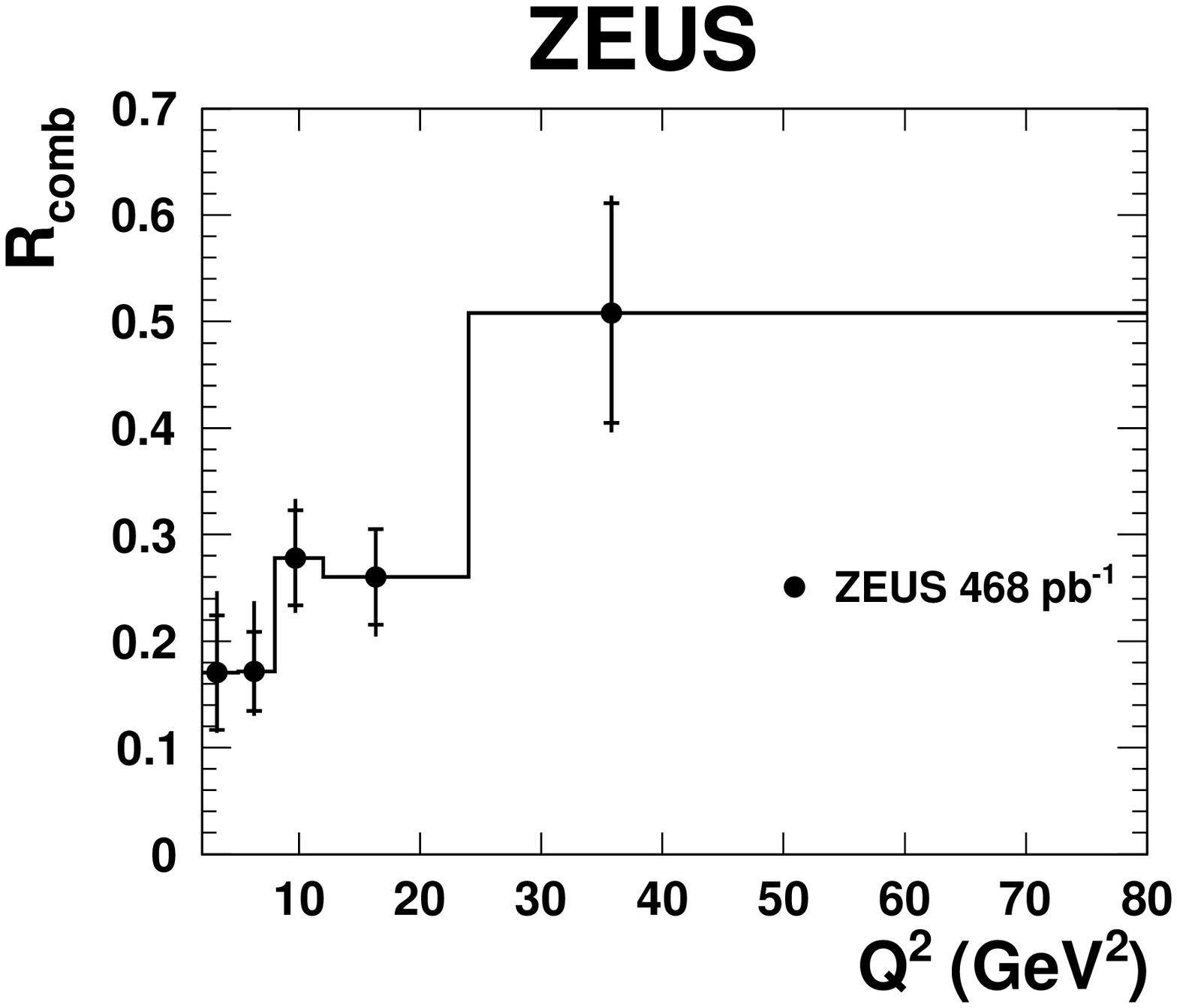}
\put(-60,55){\makebox(0,0)[tl]{ (a)}}\\
\includegraphics[width=8.5cm]{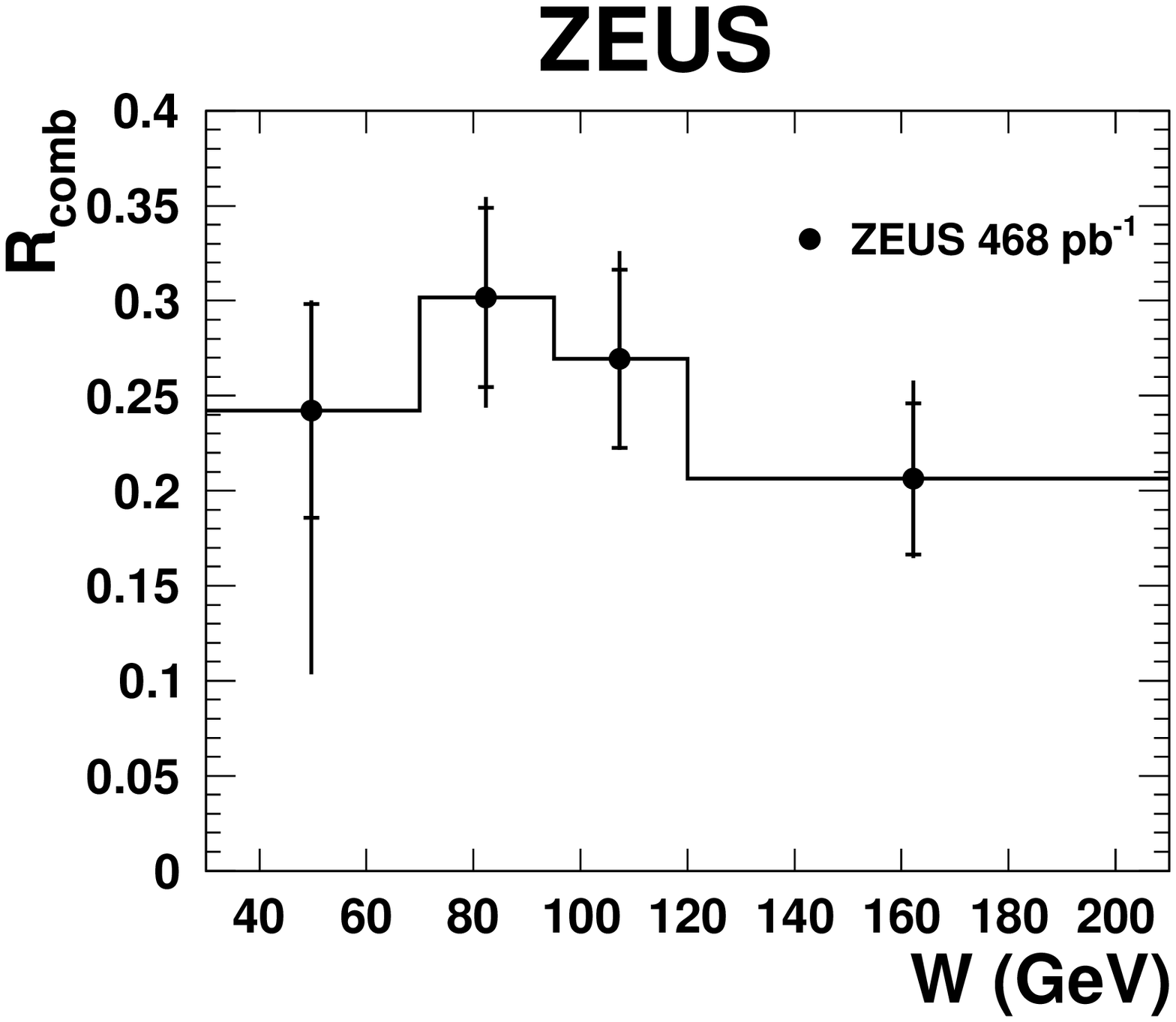}
\put(-60,55){\makebox(0,0)[tl]{(b)}}\\
\includegraphics[width=8.5cm]{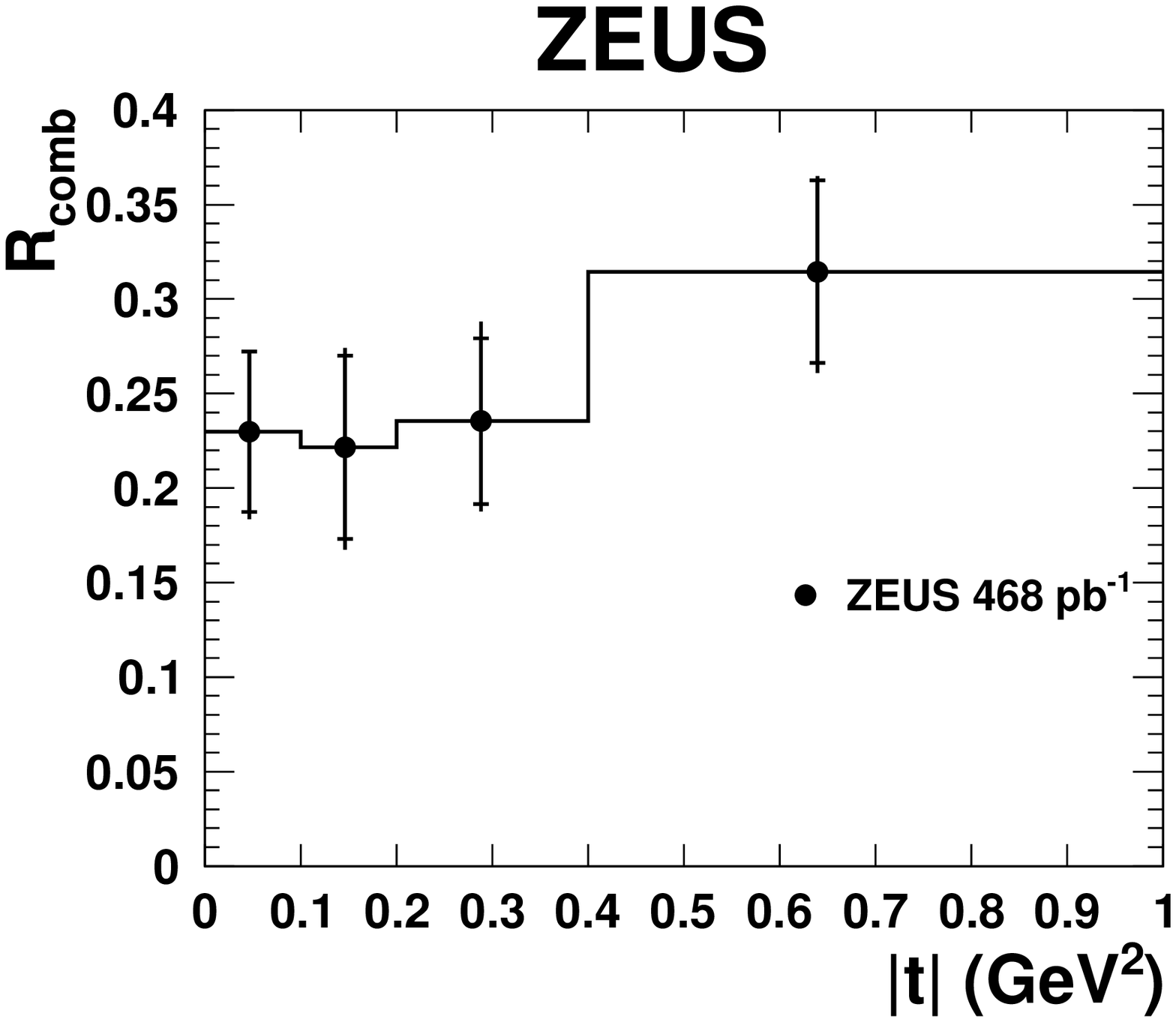}
\put(-60,55){\makebox(0,0)[tl]{ (c)}}\\
\end{center}
 \caption{Cross-section ratio $R_{\rm comb} = \sigma_{\psi(2S)}/\sigma_{J/\psi(1S)}$ for the combined $\psi(2S)$ decay modes as a function of (a) $Q^2$, (b) W and (c) |t|.
 The horizontal lines show the bin widths, and the points are plotted at the average of the reweighted $\psi (2S)$ Monte Carlo events in the corresponding bin, as recommended elsewhere\,\protect\cite{Lafferty:1994cj}.
 The inner error bars show the statistical and the outer error bars show the quadratic sum of statistical and systematic uncertainties.}
\label{fig-R-W-t}
\vfill
\end{figure}

 Figure~\ref{fig-R-W-t} shows the values of $R_{\rm comb}$ as a function of $Q^2$, $W$ and $|t|$.
 The results for the $W$ and $|t|$~dependence are shown for $5 < Q^2 < 80$\,GeV$^2$. The results for the $Q^2$ dependence also include the additional measurement for $2 < Q^2 < 5$\,GeV$^2$ from the HERA~I data.
 As a function of $W$ and $|t|$, the values of $R_{\rm comb}$ are compatible with a constant.
 For the $Q^2$ dependence, a positive slope is observed with the significance of $\approx 2.5$ standard deviations.

 As a cross check it was verified that the ratio $R_{J/\psi \,\pi \pi}$ to $R_{\mu \mu}$ is compatible with 1.
 For the entire kinematic range of the measurement a value of
 $R_{\psi(2S)} = R_{J/\psi \,\pi \pi} / R_{\mu \mu} = 1.1 \pm 0.2 \,^{+ 0.2}_{-0.1}\, \pm 0.1$ is found.
 The first error is the statistical uncertainty, the second the systematic uncertainty of the measurement and the third the uncertainty of the $\psi(2S)$ branching fractions.
 The ratio is consistent with unity for the entire kinematic region and also as a function of the kinematic variables, $Q^2$, $W$ and $|t|$, as shown in Tables~\ref{tab:Rtot} and \ref{tab:R-diff}.

\begin{figure}[p]
\vfill
\begin{center}
\includegraphics[width=17.0cm]{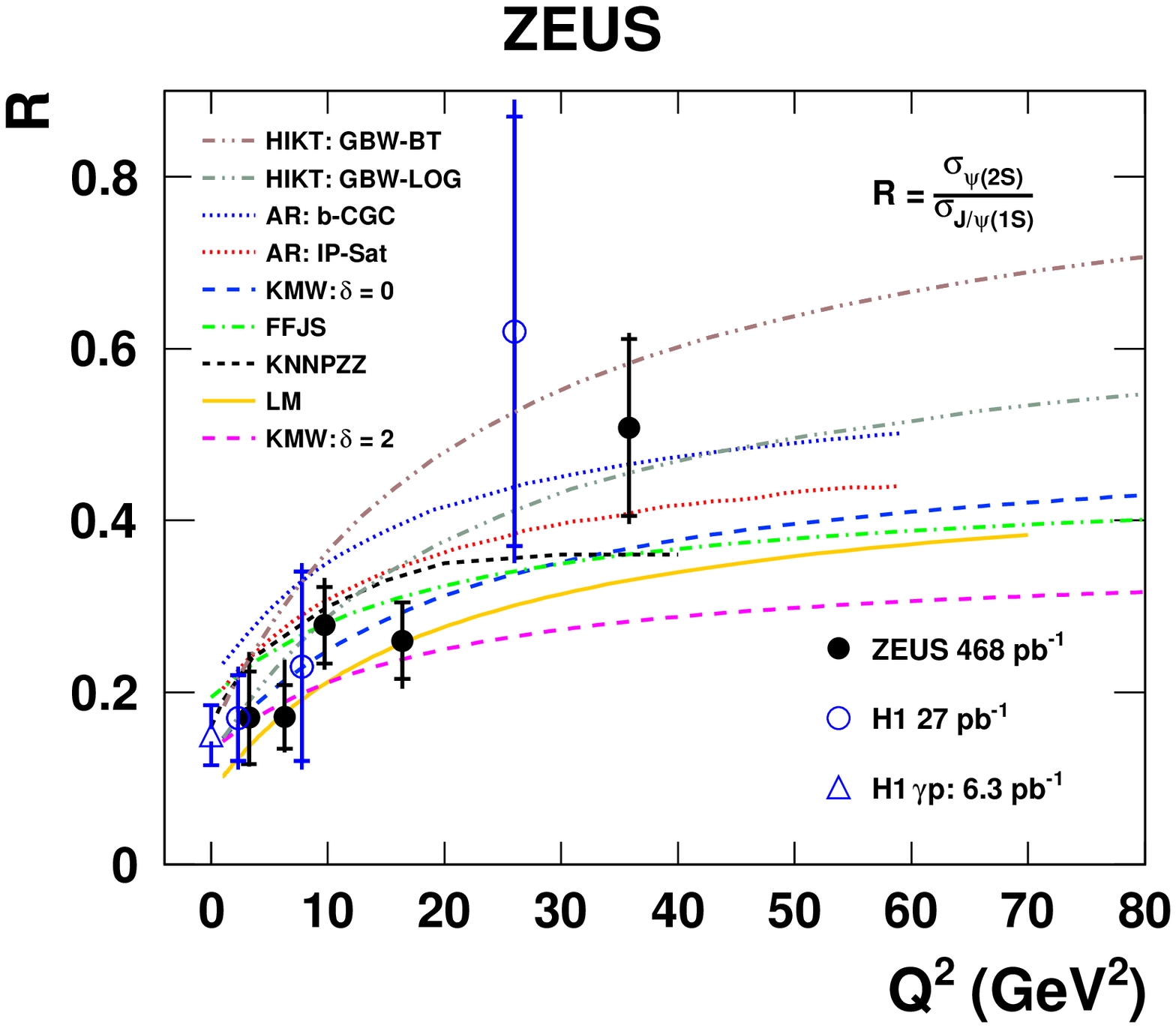}
\end{center}
 \caption{Cross-section ratio $R = \sigma_{\psi(2S)}/\sigma_{J/\psi(1S)}$ for the combined $\psi(2S)$ decay modes as a function of $Q^2$.
 The kinematic range is $30 < W < 210$\,GeV and $|t| < 1$\,GeV$^2$ at an $ep$ centre-of-mass energy of 317\,GeV for the ZEUS data with $Q^2 > 5$\,GeV$^2$ and 300\,GeV for the ZEUS data with $2 < Q^2 < 5$\,GeV$^2$.
 The ZEUS results (solid points) are shown compared to the previous H1 result (open points)~\protect\cite{epj:c10:373} measured for $25 < W < 180$\,GeV and $|t| < 1.6$\,GeV$^2$ at an $ep$ centre-of-mass energy of 300\,GeV.
 The inner error bars show the statistical and the outer error bars show the quadratic sum of statistical and systematic uncertainties.
  The ZEUS points are plotted at the average $Q^2$ of the reweighted simulated $\psi (2S)$ events with the $W$ and $t$ cuts used in the analysis, as recommended elsewhere\,\protect\cite{Lafferty:1994cj}.
 The model predictions discussed in Section~\ref{sect:Models} are shown as curves.
 The sequence of the labelling is in descending order of the $R$~value at the highest $Q^2$ of each prediction. }
\label{fig-R-Q2}
\vfill
\end{figure}

 In Fig.~\ref{fig-R-Q2}, the results are also compared to the previous H1 measurements~\cite{epj:c10:373}.
 The results are compatible. 
 The H1~collaboration has also measured $R = \sigma _{\psi(2S)}/ \sigma _{J/\psi(1S)}$ in photoproduction ($Q^2 \approx 0$), and found a value of $R =$ 0.150 $\pm$ 0.035~\cite{pl:b421:385}, which is consistent with the trend of the $Q^2$ dependence presented in this paper.
 The comparison of the results to various model predictions is presented in the Section~\ref{sect:Comparisons}.

\section{Comparison to model predictions}
      \label{sect:Comparisons}


 In this section, the cross-section ratio $R=\sigma_{\psi(2S)}/\sigma_{J/\psi(1S)}$ in DIS, presented in this paper, and the results from the H1~collaboration in DIS~\cite{epj:c10:373} and photoproduction~\cite{pl:b421:385}, are compared to model predictions from six different groups labelled by the names of the authors:
 HIKT, KNNPZZ, AR, LP, FFJS and KMW.

 \subsection{Individual models}
           \label{sect:Models}

 In order to calculate the exclusive production of vector charmonium states according to the diagram shown in Fig.~\ref{fig-diagram}, the following components need to be determined:
 \begin{itemize}
  \item the probability of finding a $c\bar{c}$-dipole of transverse size $r$ and impact parameter $b$ in the photon in the infinite momentum frame;
  \item the $c\bar{c}$-dipole scattering amplitude or cross section of the proton as a function of $r$, $b$ and $x_\mathrm{Bj} \approx (M_V + Q^2)/(W^2 + Q^2)$;
  \item the probability that the $c\bar{c}$-dipole forms the vector charmonium state $V$ in the infinite momentum frame.
 \end{itemize}
 The probability distribution of $c\bar{c}$-dipoles in the photon can be calculated in QED~\cite{Kogut:1969xa,Bjorken:1970ah}.
 For the probability that the $c\bar{c}$-dipole forms the vector charmonium state, its centre-of-mass wave function has to be boosted into the infinite momentum frame, which is done using the boosted Gaussian model~\cite{Terentev:1976jk}.
 For the evaluation of the $c\bar{c}$-dipole scattering amplitude, different QCD calculations are used.



 H\"{u}fner et al.~\cite{pr:d62:094022} (HIKT) use two phenomenological parameterisations of the $c\bar{c}$-dipole cross section, GBW\cite{GolecBiernat:1998js} and KST\cite{Kopeliovich:1999am},
 which both describe the low-$x$ inclusive DIS data from HERA.
 For the centre-of-mass wave functions of the $J/\psi(1S)$ and $\psi(2S)$, they use four different phenomenological potentials, BT, LOG, COR and POW, and $c$-quark masses between 1.48 and 1.84~GeV.
 However, only the models with $c$-quark masses around 1.5\,GeV, GBW--BT and GBW--LOG, are able to describe the cross sections of exclusive $J/\psi (1S)$ production measured at HERA.
 For the boost of the charmonium wave functions into the infinite momentum frame, they find the wave function from the Schr\"{o}dinger equation and then boost the result.
 A major progress made~\cite{pr:d62:094022} is the inclusion of the Melosh spin rotation into the boosting procedure, which enhances the $\psi(2S)$ to $J/\psi(1S)$ cross-section ratio by a factor of two to three.
 The predicted $Q^2$~dependence of the suppression of exclusive $\psi(2S)$ relative to $J/\psi(1S)$~production is caused by the node of the radial $\psi(2S)$~wave function, which leads to a destructive interference of the contributions to the production amplitude of small and large dipoles.

 The model of Kopeliovich et al.~\cite{pr:d44:3466,pl:b324:469,pl:b341:228,jetp:86:1054} (KNNPZZ) uses the running gBFKL approach for the $c\bar{c}$-dipole cross section and the diffractive slope for its $t$~dependence.
 The parameterisation of the $c\bar{c}$-dipole cross section used gives a quantitative description of the rise of the proton structure function at small $x$~values as well as of the $Q^2$ and $W$~dependence of diffractive $J/\psi(1S)$ production.
 The reason for the $Q^2$~dependence of the suppression of exclusive $\psi(2S)$ production relative to $J/\psi(1S)$~production is again related to the node of the radial $\psi(2S)$~wave function.

 Armesto and Rezaeian~\cite{pr:d90:054003} (AR) calculate the $c\bar{c}$-dipole cross section using the Impact-Parameter-dependent Color Glass Condensate model (b-CGC)~\cite{pr:d88:074016} as well as the Saturation (IP-Sat)~\cite{pr:d87:034002} dipole model, recently updated with fits to the HERA combined data~\cite{jhep:01:109,epj:c73:2311}.
 In the b-CGC model, which is restricted to the gluon sector, saturation is driven by the BFKL evolution, and its validity is therefore limited to $x_\mathrm{Bj} \lesssim 10^{-2}$.
 The IP-Sat model uses DGLAP evolution and smoothly matches the perturbative QCD limit at high values of $Q^2$.
 For the calculation of the light-cone $J/\psi(1S)$ and $\psi(2S)$ wave functions, the boosted Gaussian model and the leptonic decay widths $\Gamma_{ee}^{J/\psi(1S)}$ and $\Gamma_{ee}^{\psi(2S)}$ are used.


 Lappi and M\"{a}ntysaari~\cite{pr:c83:065202,pos:dis2014:069} (LM) use the BFKL evolution as well as the IP-Sat model to predict vector-meson production in $ep$ and electron--ion collisions in the dipole picture.
 The wave functions of the $J/\psi(1S)$ and $\psi(2S)$ have been calculated according to the procedure developed previously~\cite{Kowalski:2006hc,jhep:0906:034} and the low-$x$ inclusive HERA data have been used to constrain the $c\bar{c}$-dipole cross section.


 Fazio et al.~\cite{pr:d90:016007} (FFJS) use a two-component Pomeron model to predict the cross sections for vector-meson production.
 A normalisation factor of $f_{\psi(2S)}^{-1} = 0.45$ ensures that the value of the $\psi(2S)$ cross section is the same as for the other vector mesons at the same values of $W$, $t$ and $Q_V^2 = M_V^2+Q^2$ (i.e.~$f_{\psi(2S)}\,\sigma_{\psi(2S)} = \sigma_{J/\psi}$).

 Kowalski, Motyka and Watt\,\cite{Kowalski:2006hc} (KMW) assume the universality of the production of vector quarkonia states in the scaling variable $Q_V^2$ in their calculation of $R$.
 With the assumptions that the $c\bar{c} \to V$ transition is proportional to the leptonic decay width $\Gamma_{ee}^V$ and that the interaction is mediated by two-gluon exchange and therefore proportional to $\big(\alpha_s(Q_V)\,x_\mathrm{Bj}\,g(x_\mathrm{Bj},Q_V^2)\big)^2$, $R$ is given in the leading-logarithmic approximation~\cite{Ryskin:1992ui,Brodsky:1994kf,Jones:2013pga} by
\begin{equation}
\label{ratioparam}
 R = \left( {\alpha_s(Q_{\psi(2S)}) \over \alpha_s (Q _{J/\psi(1S)})} \right)^2\, {\Gamma_{\psi(2S)} M_{\psi(2S)} ^{1-\delta} \over \Gamma_{J/\psi(1S)} M_{J/\psi(1S)} ^{1-\delta} }\, \left({Q_{\psi(2S)} \over Q_{J/\psi(1S)}} \right)^{-6-4\bar\lambda + \delta}.
\end{equation}
 The running strong coupling constant, $\alpha _s (Q)$, and the gluon density, $g(x, Q^2)$, are evaluated at $Q_V$ and $x_\mathrm{Bj}$.
 For small $x_{\mathrm{Bj}}$~values, the gluon density can be parameterised as $x_{\mathrm{Bj}} \, g(x_{\mathrm{Bj}},Q_V^2) \propto x_{\mathrm{Bj}}^{-\lambda(Q_V)}$ with $\lambda(Q_V) \simeq \bar \lambda = 0.25$ in the $Q_V$~region of this measurement.
 The parameter $\delta$ depends on the choice of the charmonium wave functions.
 For the non-relativistic  wave functions $\delta=0$~\cite{Ryskin:1992ui, Brodsky:1994kf, Jones:2013pga} and for the relativistic boosted Gaussian model $\delta \approx 2$~\cite{Kowalski:2006hc}. 
 Note that the $Q^2$ dependence of the ratio $R$ in this approach is driven by kinematic factors and not by the shapes and nodes of the charmonia wave functions.

 \subsection{Comparison of models and data}
           \label{sect:Data_comparison}

 In the kinematic region of the measurement, all models predict only a weak $W$ and $|t|$~dependence of $R$, consistent with the data, as shown in Fig.~\ref{fig-R-W-t}.
 Therefore, only the comparison of the model calculations with the measurements as a function of $Q^2$ is presented in Fig.~\ref{fig-R-Q2}.
 It can be seen that all models predict an increase of $R$ with $Q^2$, which is also observed experimentally.
 In the following discussion, the models are discussed in the sequence from higher to lower predicted $R$ values at high $Q^2$.

 From the HIKT calculations, the results for $R$ for the two charmonium potentials BT and LOG with $c$-quark mass around 1.5~GeV and the GBW model for the $c\bar{c}$-dipole cross section are shown.
 They provide a good description of the energy dependence of exclusive $J/\psi(1S)$ production for centre-of-mass energies between 10 and 300~GeV, which is not the case for the potentials COR and POW with $c$-quark masses of about 1.8~GeV.
 The difference of the results when using the KST dipole cross sections are small.
 For $Q^2$ values below 24\,GeV$^2$, the predicted $R$~values for the BT model are significantly larger than the measured values. 
 For the LOG model, the predicted values are closer to the data.
 
 For the AR\,calculations, the results for the b-CGC and IP-Sat models of the dipole cross sections are shown in the figure.
 The b-CGC prediction for $R$ is about 20\,\% higher than that for IP-Sat. 
 For $Q^2$~values below 24\,GeV$^2$, the IP-Sat model gives a better description of the data.
 
 The KMW\,model with $\delta = 0$ provides a good description of the observed $Q^2$\,dependence of $R$, whereas the prediction for $\delta = 2$ is below the $R$~value measured for $Q^2 > 24$\,GeV$^2$, which however has a large experimental uncertainty.
 
 The predictions of the models FFJS, KNNPZZ and LM, in spite of differences in the values of $R$ at low $Q^2$~values, also provide good descriptions of the measurements.
 
 Some discrimination of the different models is possible, although a large spread in the predictions indicates that the uncertainty on the theory is large.

 \section{Summary}
      \label{sect:Summary}
 The cross-section ratio $R = \sigma _{\psi(2S)}/ \sigma _{J/\psi(1S)}$ in exclusive electroproduction was measured with the ZEUS experiment at HERA in the kinematic range $2 <Q^2 <80$\,GeV$^2$, $30 <W <210$\,GeV and $|t| < 1$\,GeV$^2$, with an integrated luminosity of 468\,pb$^{-1}$.
 The decay channels used were $\mu^+ \mu^-$ and $ J/\psi(1S) \,\pi^+ \pi^-$ for the $\psi(2S)$ and $\mu^+ \mu^-$ for the $ J/\psi (1S)$.
 The cross-section ratio has been determined as a function of $Q^2$, $W$ and $|t|$.
 The results are consistent with a constant value of $R$ as a function of $W$ and $t$, but show a tendency to increase with increasing $Q^2$.
 A number of model calculations are compared to the measured $Q^2$ dependence of $R$.

\section*{Acknowledgements}
\label{sec-ack}

 We appreciate the contributions to the construction, maintenance and operation of the ZEUS detector of many people who are not listed as authors.
 The HERA machine group and the DESY computing staff are especially acknowledged for their success in providing excellent operation of the collider and the data-analysis environment.
 We thank the DESY directorate for their strong support and encouragement.
 We also thank Y.\,Ivanov, L.\,Jenkovski, B.\,Kopeliovich, L.\,Motyka, A.\,Rezaeian and A.\,Salii for interesting discussions and for providing the results of their calculations.

\clearpage

{
\ifzeusbst
  \ifzmcite
     \bibliographystyle{./BiBTeX/bst/l4z_default3}
  \else
     \bibliographystyle{./BiBTeX/bst/l4z_default3_nomcite}
  \fi
\fi
\ifzdrftbst
  \ifzmcite
    \bibliographystyle{./BiBTeX/bst/l4z_draft3}
  \else
    \bibliographystyle{./BiBTeX/bst/l4z_draft3_nomcite}
  \fi
\fi
\ifzbstepj
  \ifzmcite
    \bibliographystyle{./BiBTeX/bst/l4z_epj3}
  \else
    \bibliographystyle{./BiBTeX/bst/l4z_epj3_nomcite}
  \fi
\fi
\ifzbstjhep
  \ifzmcite
    \bibliographystyle{./BiBTeX/bst/l4z_jhep3}
  \else
    \bibliographystyle{./BiBTeX/bst/l4z_jhep3_nomcite}
  \fi
\fi
\ifzbstnp
  \ifzmcite
    \bibliographystyle{./BiBTeX/bst/l4z_np3}
  \else
    \bibliographystyle{./BiBTeX/bst/l4z_np3_nomcite}
  \fi
\fi
\ifzbstpl
  \ifzmcite
    \bibliographystyle{./BiBTeX/bst/l4z_pl3}
  \else
    \bibliographystyle{./BiBTeX/bst/l4z_pl3_nomcite}
  \fi
\fi
{\raggedright
\bibliography{./BiBTeX/bib/l4z_zeus.bib,%
              ./BiBTeX/bib/l4z_h1.bib,%
              ./BiBTeX/bib/l4z_articles.bib,%
              ./BiBTeX/bib/l4z_books.bib,%
              ./BiBTeX/bib/l4z_conferences.bib,%
              ./BiBTeX/bib/l4z_misc.bib,%
              ./BiBTeX/bib/l4z_preprints.bib}}
}
\vfill\eject

\setcounter{table}{0}

\begin{table}
 \begin{center}
  \caption{
  Measured cross-section ratios $\sigma_{\psi(2S)}/\sigma_{J/\psi(1S)}$:
   $R_{J/\psi \pi \pi}$ for the decay $\psi (2S) \to J/\psi(1S)\,\pi^{+} \pi^{-}$,
   $R_{\mu \mu}$ for the decay $\psi (2S) \to \mu^{+} \mu^{-}$,
   and $R_{\rm comb}$ for the two decay modes combined,
  for the kinematic range $5 < Q^2 < 80$\,\rm{GeV}$^2$, $30 < W < 210$\,\rm{GeV}
  \textit{and} $|t| < 1\,\rm{GeV}^2$
  \textit{at an $ep$ centre-of-mass energy of} $317\,\rm{GeV}$.
  \textit{Also shown is $R_{\psi (2S)} = R_{J/\psi \pi \pi} / R_{\mu \mu}$.}
   \textit{Statistical and systematic uncertainties are given.}
 \label{tab:Rtot}}
\vspace{0.5cm}
\begin{tabular}{|l|c|}
\hline
$ R_{J/\psi \pi \pi}$& $0.26 \pm 0.03 _{- 0.01 } ^{+ 0.01 }$ \\
\hline
$ R_{\mu \mu}$              & $0.24 \pm 0.05 _{- 0.03 } ^{+ 0.02 }$\\
\hline
$ R_{\rm comb}$                                & $0.26 \pm 0.02 _{- 0.01 } ^{+ 0.01 }$\\
\hline
\hline
$R_{\psi(2S)}$                      & $1.1 \pm 0.2 _{- 0.1 } ^{+ 0.2 }$\\
\hline
\end{tabular}
\end{center}
\end{table}
\begin{landscape}
\begin{table}
\begin{center}
\caption{
 Measured cross-section ratios $\sigma_{\psi(2S)}/\sigma_{J/\psi(1S)}$:
  $R_{J/\psi \pi \pi}$ for the decay $\psi (2S) \to J/\psi(1S)\,\pi^{+} \pi^{-}$,
  $R_{\mu \mu}$ for the decay $\psi (2S) \to \mu^{+} \mu^{-}$,
  and $R_{\rm comb}$ for the two decay modes combined, as a function of $Q^2$, $W$ and $|t|$.
 Also shown is $R_{\psi (2S)} = R_{J/\psi \pi \pi} / R_{\mu \mu}$.
 The ratios as a function of W are measured in the range $5 < Q^2 < 80$\,\rm{GeV}$^2$ \textit{and}
 $|t| < 1\,\rm{GeV}^2$,
 \textit{as a function of} $Q^2$
 \textit{in the range} $30 < W < 210$\,\rm{GeV}
 \textit{and} $|t| < 1\,\rm{GeV}^2$,
 \textit{and as a function of $|t|$ in the range} $5 < Q^2 < 80$\,\rm{GeV}$^2$
 \textit{and} $30 < W < 210$\,\rm{GeV}.
 \textit{All results are quoted for an $ep$ centre-of-mass energy of} $317\,\rm{GeV}$,
 \textit{except for the bin} $2 < Q^2 < 5$\,\rm{GeV}$^2$,
 \textit{which refers to} $300\,\rm{GeV}$.
 \textit{Statistical and systematic uncertainties are given.}
\label{tab:R-diff}
}
\vspace{3mm}
\begin{tabular}{|c|c|c|c||c|}

\hline
$Q^2$ $(\rm GeV^{2})$   &   $R_{J/\psi\pi\pi}$  &  $R_{\mu\mu}$   &  $R_{\rm comb}$   &  $R_{\psi(2S)}$\\
\hline
$2 - 5$ & $0.21 \pm 0.07 _{- 0.03 } ^{+ 0.04 }$ & $0.10 \pm 0.09 _{- 0.09 } ^{+ 0.09 }$ & $0.17 \pm 0.05 _{- 0.02 } ^{+ 0.05}$ & -- \\
 $ 5 - 8$ & $0.19 \pm 0.05 _{- 0.02 } ^{+ 0.02 }$ & $0.13 \pm 0.06 _{- 0.03 } ^{+ 0.12 }$ & $0.17 \pm 0.04 _{- 0.02 } ^{+ 0.05 } $ & $1.5 \pm 0.8 _{- 0.7 } ^{+ 0.4 } $\\
 $ 8 - 12$ & $0.27 \pm 0.05 _{- 0.01 } ^{+ 0.06 }$ & $0.29 \pm 0.08 _{- 0.08 } ^{+ 0.03 }$ & $0.28 \pm 0.05 _{- 0.03 } ^{+ 0.03 } $ & $0.9 \pm 0.3 _{- 0.1 } ^{+ 0.4 } $\\
 $ 12 - 24$ & $0.27 \pm 0.05 _{- 0.03 } ^{+ 0.04 }$ & $0.24 \pm 0.08 _{- 0.08 } ^{+ 0.01 }$ & $0.26 \pm 0.05 _{- 0.03 } ^{+ 0.01 } $ & $1.1 \pm 0.4 _{- 0.1 } ^{+ 0.6 } $\\
 $ 24 - 80$ & $0.56 \pm 0.13 _{- 0.09 } ^{+ 0.04 }$ & $0.42 \pm 0.17 _{- 0.04 } ^{+ 0.12 }$ & $0.51 \pm 0.10 _{- 0.04 } ^{+ 0.04 } $ & $1.3 \pm 0.6 _{- 0.6 } ^{+ 0.3 } $\\

\hline
\hline
$W$ $(\rm GeV)$   &   $R_{J/\psi\pi\pi}$  &  $R_{\mu\mu}$   &  $R_{\rm comb}$   &  $R_{\psi(2S)}$\\
\hline
 $ 30 - 70$ & $0.24 \pm 0.07 _{- 0.13 } ^{+ 0.01 }$ & $0.24 \pm 0.10 _{- 0.14 } ^{+ 0.03 }$ & $0.24 \pm 0.06 _{- 0.13 } ^{+ 0.01 } $ & $1.0 \pm 0.5 _{- 0.2 } ^{+ 0.5 } $\\
 $ 70 - 95$ & $0.30 \pm 0.06 _{- 0.04} ^{+ 0.01 }$ & $0.31 \pm 0.09 _{- 0.03 } ^{+ 0.09 }$ & $0.30 \pm 0.05 _{- 0.03 } ^{+ 0.02 } $ & $1.0 \pm 0.3 _{- 0.2 } ^{+ 0.1 } $\\
 $ 95 - 120$ & $0.28 \pm 0.06 _{- 0.01 } ^{+ 0.05 }$ & $0.24 \pm 0.08 _{- 0.05 } ^{+ 0.04 }$ & $0.27 \pm 0.05 _{- 0.01 } ^{+ 0.03 } $ & $1.2 \pm 0.5 _{- 0.2 } ^{+ 0.5 } $\\
 $ 120 - 210$ & $0.22 \pm 0.05 _{- 0.01 } ^{+ 0.07 }$ & $0.17 \pm 0.07 _{- 0.05 } ^{+ 0.02 }$ & $0.21 \pm 0.04 _{- 0.01 } ^{+ 0.03 } $ & $1.3 \pm 0.6 _{- 0.2 } ^{+ 0.7 } $\\
\hline
\hline
$|t|$ $(\rm GeV^{2})$   &   $R_{J/\psi\pi\pi}$  &  $R_{\mu\mu}$   &  $R_{\rm comb}$   &  $R_{\psi(2S)}$\\
\hline
 $ 0 - 0.1$ & $0.23 \pm 0.05 _{- 0.02 } ^{+ 0.02 }$ & $0.23 \pm 0.09 _{- 0.05 } ^{+ 0.04 }$ & $0.23 \pm 0.04 _{- 0.02 } ^{+ 0.01 } $ & $1.0 \pm 0.4 _{- 0.2 } ^{+ 0.3 } $\\
 $ 0.1 - 0.2$ & $0.22 \pm 0.06 _{- 0.03 } ^{+ 0.02 }$ & $0.23 \pm 0.09 _{- 0.06 } ^{+ 0.02 }$ & $0.22 \pm 0.05 _{- 0.02 } ^{+ 0.02 } $ & $0.9 \pm 0.4 _{- 0.2 } ^{+ 0.5 } $\\
 $ 0.2 - 0.4$ & $0.27 \pm 0.06 _{- 0.01 } ^{+ 0.06 }$ & $0.18 \pm 0.07 _{- 0.06 } ^{+ 0.05 }$ & $0.24 \pm 0.04 _{- 0.02 } ^{+ 0.03 } $ & $1.5 \pm 0.6 _{- 0.2 } ^{+ 0.5 } $\\
 $ 0.4 - 1$ & $0.32 \pm 0.06 _{- 0.03 } ^{+ 0.05 }$ & $0.30 \pm 0.08 _{- 0.05 } ^{+ 0.02 }$ & $0.32 \pm 0.05 _{- 0.02 } ^{+ 0.01 } $ & $1.1 \pm 0.3 _{- 0.1 } ^{+ 0.3 } $\\
\hline
\end{tabular}
\end{center}
\end{table}
\end{landscape}

%
%
\end{document}